\documentclass[twocolumn,trackchanges]{aastex631}
\usepackage{amsmath}
\usepackage{amssymb}
\usepackage{amsthm}
\usepackage{natbib,aasdefs,url,bm}
\usepackage{array}
\usepackage{float}
\usepackage{graphicx}
\usepackage{subfigure}
\hypersetup{linkcolor=red,citecolor=NavyBlue,filecolor=cyan,urlcolor=purple}

\def\be{\begin{equation}}
\def\ee{\end{equation}}
\def\ba{\begin{eqnarray}}
\def\ea{\end{eqnarray}}

\def\aap{\ {A\&A}\ }

\def\apjl{\ {ApJL}\ }
\def\apjs{\ {ApJS}\ }

\submitjournal{ApJ}
\accepted{April 2, 2025}
\shorttitle{Neutrino signatures from pulsar remnants of BNS mergers}
\shortauthors{Mukhopadhyay et al.}

\begin{document}
\title{
High-energy neutrino signatures from pulsar remnants of binary neutron-star mergers: coincident detection prospects with gravitational waves
}
\correspondingauthor{Mainak Mukhopadhyay}
\email{mkm7190@psu.edu}
\author[0000-0002-2109-5315]{Mainak Mukhopadhyay}
\affiliation{Department of Physics; Department of Astronomy \& Astrophysics; Center for Multimessenger Astrophysics, Institute for Gravitation and the Cosmos, The Pennsylvania State University, University Park, PA 16802, USA}
\author[0000-0003-2579-7266]{Shigeo S. Kimura}
\affiliation{Frontier Research Institute for Interdisciplinary Sciences; Astronomical Institute, Graduate School of Science, Tohoku University, Sendai 980-8578, Japan}
\author[0000-0002-4670-7509]{Brian D. Metzger}
\affiliation{Department of Physics and Columbia Astrophysics Laboratory, Columbia University, New York, NY 10027, USA}
\affiliation{Center for Computational Astrophysics, Flatiron Institute, 162 5th Ave, New York, NY 10010, USA}
\begin{abstract}
Binary neutron-star (BNS) mergers are accompanied by multi-messenger emissions, including gravitational wave (GW), neutrino, and electromagnetic signals. Some fraction of BNS mergers may result in a rapidly spinning magnetar as a remnant, which can enhance both the EM and neutrino emissions. 
In this study, we model the neutrino emissions from such systems and discuss the prospects for detecting the neutrinos coincident with GW signatures.
We consider a scenario where a magnetar remnant drives a pulsar wind using its spin energy. The wind interacts with the surrounding kilonova ejecta, forming a nebula filled with non-thermal photons. Ions and nuclei extracted from the magnetar's surface can be accelerated in the polar-cap and the termination-shock regions. We investigate the neutrino fluences resulting from photomeson interactions, where accelerated CR protons interact with the photons in the nebula. Our findings indicate that the peak neutrino fluence is $\sim 10^{-2}\rm GeV~cm^{-2}$ for a source at $40$ Mpc, which is reached approximately $\mathcal{O}( 1-10\ {\rm days})$ post merger. Finally, we examine the potential for GW-triggered stacking searches with IceCube-Gen2 using next-generation GW detectors such as the Cosmic Explorer (CE) and the Einstein Telescope (ET). We conclude that, assuming an optimistic neutrino emission model, a combination of CE+ET would offer a high probability of neutrino detection from these sources within an operational timescale of $\sim 20$ years. In case of non-detection, $2 \sigma$ level constraints on model parameters can be established within similar joint operation timescales.
\end{abstract}
\section{Introduction}
\label{sec:intro}
Binary neutron star (BNS) mergers are promising sources of multi-messenger emissions~\citep{1989Natur.340..126E,Rosswog:2015nja,Baiotti:2016qnr,Blinnikov:2018boq,Metzger2019,Nakar:2019fza,Ascenzi:2020xqi,Ruiz:2021gsv}. During the inspiral phase of two neutron stars (NSs), gravitational waves (GWs) extract the binary angular momentum, ultimately leading to a merger event. A portion of the enormous binding energy of the BNS system is released in GWs, rendering these events ideal GW sources~\citep{Baiotti:2019sew,Dietrich:2020eud}. A BNS merger also creates a central compact object remnant surrounded by an expanding ejecta shell. The remnant can launch relativistic jets, which can power observable EM signatures in multi-wavelength bands~\citep{Fernandez:2015use,Combi:2023yav} and can also be a promising site for high-energy neutrino production~\citep{Waxman:1997ti,Murase:2019tjj,Kimura:2022zyg}. 

The discovery of GW170817 in GW~\citep{LIGOScientific:2017vwq} and EM~\citep{Goldstein:2017mmi,LIGOScientific:2017zic,LIGOScientific:2017ync,DES:2017kbs,Coulter:2017wya,J-GEM:2017tyx,Valenti:2017ngx,Lipunov:2017dwd,Chornock:2017sdf,Drout:2017ijr,Haggard:2017qne,Hallinan:2017woc,Kilpatrick:2017mhz,Margutti:2017cjl,Pian:2017gtc,Savchenko:2017ffs,Shappee:2017zly,Smartt:2017fuw,Troja:2017nqp,DAvanzo:2018zyz,Ghirlanda:2018uyx} channels further established the above fact. However, no neutrinos were detected from the event~\citep{ANTARES:2017bia,Super-Kamiokande:2018dbf}. This leads to a couple of intriguing and timely questions, particularly in the light of planned next-generation neutrino and GW detectors: (i) what are the plausible templates and light curves of neutrino emission from such sources? and (ii) given the large horizon distances of the next generation of GW detectors, how can information extracted from the GW signal be used to perform informed searches for high-energy neutrinos from the resulting remnants?

In this work, we attempt to address both of these questions. For (i), we model a magnetar-nebula-ejecta system and study its dynamical evolution. Within this time-dependent environment we explore CR particle acceleration and neutrino production to predict the emission from such sources. For (ii), we use the technique developed in~\cite{Mukhopadhyay:2023niv,Mukhopadhyay:2024lwq}, to study the prospects of performing GW triggered stacking searches of mergers detected by Cosmic Explorer (CE)~\citep{Reitze:2019iox} and Einstein Telescope (ET)~\citep{Maggiore:2019uih}, for high-energy neutrinos resulting from our model, with IceCube-Gen2~\citep{IceCube-Gen2:2020qha,Gen2_TDR}.

Remnants spanning a wide variety of properties can form from BNS mergers. The outcome depends on the masses and spins of the NSs, as well as the nuclear-density equation of state (EOS), which is currently not well constrained. In one set of scenarios, the remnant collapses almost immediately to form a black hole (BH), whereas in another it forms a rapidly spinning NS remnant. In the latter case, there are three possible outcomes: a) a ``hyper-massive'' NS, which still collapses relatively promptly collapses to a BH, likely within a second or less of the merger; b) a ``supra-massive'' NS, which is supported by rapid solid-body rotation and only collapses to a BH over longer timescales (e.g., $\mathcal{O}(10^5 \rm s)$); or c) an infinitely stable NS that never collapses to a BH (see~\citealt{2019ARNPS..69...41S,Sarin:2020gxb} and references therein for detailed reviews)\footnote{Note that these outcomes are not distinct and form a spectrum~\citep{Radice:2020ddv}. The existence of a sharp line at $1.2\ \rm M_{\rm TOV}$ (where $M_{\rm TOV}$ is the maximum non-rotating neutron star mass that is supported by the equation of state) separating the various outcomes is unclear~\citep{Margalit:2022rde}.}. In this work, we focus on scenario (c), where we have a stable differentially spinning NS (or pulsar) with a strong dipolar magnetic field -- a \emph{magnetar}\footnote{In the current context, we have a rotationally powered neutron star with a high magnetic field, for which the word magnetar is used colloquially. These newly born millisecond magnetars have also been invoked to explain recently discovered class of fast X-ray transients~\citep{Sun:2019jaz,Quirola-Vasquez:2023eye}. In general, the term magnetar also refers to a class of young isolated neutron stars that have a variety of bursting activity in the electromagnetic spectrum (in particular X-rays and soft $\gamma$-rays), where the emissions are powered by decay of the enormous internal magnetic fields (see~\citealt{Kaspi:2017fwg} for a review).}.

A magnetar provides multiple sites for particle acceleration to ultra-high energies~\citep{Gunn:1969ej,Piro:2016jaq}. The ions or nuclei extracted from the surface of the magnetar can be accelerated at the polar cap region~\citep{Blasi:2000xm,Arons:2002yj,Fang:2017tla}, in the magnetar wind region~\citep{Arons:2002yj}, in the equatorial current sheet~\citep{Murase:2009pg,Philippov:2013tpa,Chen:2014dva,Cerutti:2014ysa,Cerutti:2016ttn} or at the termination shock (TS) region~\citep{Lemoine:2014ala,Kotera:2015pya}. The accelerated cosmic rays (CRs) can interact with the photons in a wind-powered nebula consisting of copious $e^+-e^-$ pairs. The nebular region provides a rich source of electromagnetic (EM) emissions which are then attenuated by the surrounding kilonova ejecta. The interaction of the accelerated CRs with the photons can produce neutral and charged pions, leading to additional EM signatures~\citep{Murase:2014bfa} and high-energy neutrinos~\citep{Murase:2009pg,Gao:2013rxa,Fang:2013vla,Fang:2017tla,Fang:2018hjp}. Besides pion decays, the decay of kaons and charmed hadrons can also contribute to ultra-high energy neutrino fluences~\citep{Carpio:2020wzg}.

The paper is organized as follows. In Section~\ref{sec:model} we discuss the various ingredients and present a general overview of our model. The dynamics of the magnetar-nebula-ejecta system is studied in Section~\ref{subsec:dyn}. The details related to CR proton acceleration and subsequent neutrino production is studied in Section~\ref{sec:nu_prod}. The main results of this work are presented in Section~\ref{sec:res}. We summarize in Section~\ref{sec:summary} and conclude with discussing the implications of our work in Section~\ref{sec:disc}. We denote the comoving quantities with a prime, unless otherwise stated. For a given particle species $i$, we use $\varepsilon_i$ to denote its individual energies and $E_i$ to denote its total energy. The time elapsed since the merger is given by $t$ in the rest frame of the central engine.
\section{Overview}
\label{sec:model}
In this section, we overview our model for the neutrino emission powered by a magnetar remnant of BNS merger within a one-zone framework. In particular, we will focus on the various physical quantities of interest along with the emission zones and mechanisms relevant to high-energy neutrino production from such a source.  

Let us first summarize the time evolution of the model. A BNS merger leaves a central compact object remnant along with an ejecta primarily composed of $r$-process nuclei. We consider a relatively low-mass merger, which leaves an indefinitely stable, rapidly-rotating magnetar. The magnetar eventually loses its spin energy as a result of magnetically-driven (pulsar-like) winds. The wind interacts with the ejecta, leading to the formation of a nebular region consisting of non-thermal photons, a part of which are eventually re-processed to thermal photons. CR protons can be accelerated in such a system at the magnetosphere of the magnetar and at the termination shock (TS) region of the wind. This forms an ideal site for high-energy neutrino production as a result of the interaction between the protons and the nebular photons.

\begin{figure}
\includegraphics[width=0.45\textwidth]{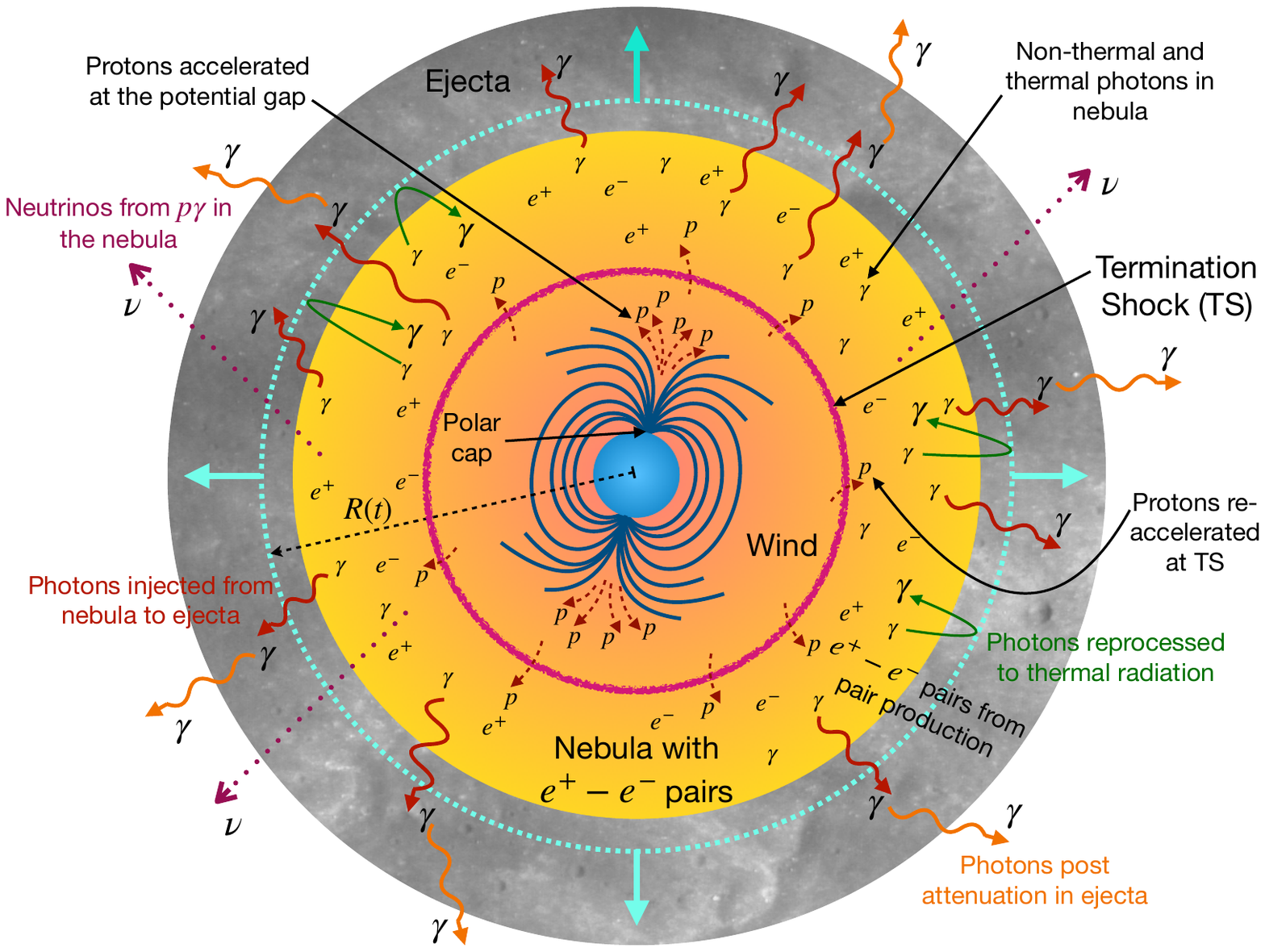}
\caption{\label{fig:model} The model of a stable millisecond magnetar we consider in this work. We show the spinning magnetar remnant from the BNS merger in blue along with the magnetic field lines in dark blue. The termination shock (TS) is shown as a thick red line which encloses the magnetar wind region shaded in orange-yellow. The nebular region is shown in yellow. The outward expanding ejecta is shown as the grey shell. We also highlight the site of proton acceleration (dashed brown arrows) at the polar cap and at the TS, the nebular region with $e^+ - e^-$ pairs, non-thermal, and thermal photons, relevant for neutrino and EM emissions.
}
\end{figure}

A schematic (time snapshot) of the model for the system is shown in Figure~\ref{fig:model}.
The magnetar is spinning with a millisecond rotation period, as determined by the large angular momentum of the original binary at the time of merger, and a strong dipolar magnetic field of $\mathcal{O}(10^{13} {\rm G}) - \mathcal{O}(10^{15} {\rm G})$. The magnetar is shown as the blue blob in the figure along with the magnetic field lines in dark blue. Any debris disk formed around the magnetar is promptly accreted or becomes unbound to form$-$along with any dynamical ejecta$-$an outward expanding ejecta shell \citep{Metzger:2014ila}, shown as the grey shell. The magnetic spin-down of the magnetar acts as an energy source which aids in forming a hot nebula behind the expanding ejecta. The spin-down energy is deposited behind the expanding ejecta using shocks or magnetic reconnection. The magnetar wind is composed of copious $e^+ - e^-$ pairs, extracted from the magnetosphere via $B-\gamma$ and $\gamma-\gamma$ interactions. The magnetar wind region is denoted by a shade of orange-yellow. The boundary between the wind and the nebular region shown in yellow is characterized by the termination shock (TS). The TS is shown as a thick red line. The nebular region consists of thermal and non-thermal photons as a result of pair annihilation of the $e^+ - e^-$ from the magnetar wind. It also consists of a sea of $e^+ - e^-$ pairs as a result of pair production from the photons, both of which are also shown in the figure.

We consider two main sites of particle acceleration (in particular protons) - the polar cap region due to a potential gap and the termination shock region. The particle acceleration sites of the protons ($p$) are marked by dashed brown arrows in the figure. The protons from the polar cap region are accelerated due to the potential gap. A fraction of them can be re-accelerated at the TS. The protons then interact with the photons in the nebular region to produce neutrinos through the photomeson ($p-\gamma$) process. The neutrinos resulting from this process are shown with dotted maroon arrows. We will mainly discuss neutrino production in this work and investigate the aspects of GW-triggered searches for neutrinos from magnetar remnants of BNS mergers. The EM signatures will be discussed in a separate work~\citep{Mukhopadhyay:2025tvz}. 

The nebula is surrounded by the ejecta, shown in grey in the figure. We consider an approximate effective boundary between the nebular and the ejecta region denoted by the dashed light blue line in the figure. The nebula and the ejecta expand outwards with some Lorentz factor $\Gamma_{\rm ej} = 1/\big( \sqrt{1-\beta_{\rm ej}^2} \big)$ and the radial distance from the center is given by $R(t)$. The ejecta region can absorb the non-thermal photons from the nebular region and reprocess them into thermal photons. This depends on the albedo and opacity of the ejecta which in turn depends on its ionization state. The photons that are reprocessed from non-thermal to thermal photons are denoted by solid green arrows in the figure. The photons that diffuse from the nebula to the ejecta are shown using curly dark red arrows. The absorbed photons then suffer attenuation based on the composition of the ejecta. The photons emerging out of the ejecta are shown as curly orange arrows.

\section{Dynamics}
\label{subsec:dyn}
%
The co-evolution of the magnetar, wind, nebula, and ejecta system plays an important role in how the magnetar's spin-down powers the nebular emission. The dynamics of similar pulsar wind nebulae (PWNe) systems have been explored in~\cite{Yu:2013kra,Metzger:2013cha, Metzger:2013kia,Siegel:2015swa,Siegel:2015twa,Fang:2017tla, Sun:2019jaz}. In particular, high-energy neutrino emissions from magnetar remnants of BNS mergers were explored in~\cite{Fang:2017tla}. We improve the previous models in various ways including but not limited to adding relativistic corrections, solving the CR proton transport equations in the steady state limit, including curvature losses during protons acceleration at the polar cap, considering two separate acceleration sites, solving for the albedo of the nebula-ejecta boundary. However, this leads to our results being less optimistic than what was presented in earlier works. 

We begin by discussing the time-evolution of the thermal ($E_{\rm th}$), non-thermal ($E_{\rm nth}$), and magnetic energy ($E_B$) in the nebula. The spin down energy of the magnetar ($E_{\rm sd}$) is distributed amongst these energies. The evolution of these energies are given by the following equations:
\begin{equation}
\label{eq:nth}
\frac{d E_{\rm nth}}{dt} = L_{\rm sd} - \frac{E_{\rm nth}}{R} \frac{d R}{dt} - \frac{E_{\rm nth}}{t^{\rm neb}_{\rm diff}}\,,
\end{equation}
\begin{equation}
\label{eq:th}
\frac{d E_{\rm th}}{dt} = \big( 1 - \mathcal{A} \big) \frac{E_{\rm nth}}{t^{\rm neb}_{\rm diff}} - \frac{E_{\rm th}}{R} \frac{d R}{dt} - \frac{E_{\rm th}}{t^{\rm ej}_{\rm diff}} + Q^{\rm heat}_{\rm rp}\,,
\end{equation}
\begin{equation}
\label{eq:eb}
\frac{d E_{B}}{dt} = \varepsilon_B L_{\rm sd} - \frac{E_B}{R} \frac{dR}{dt}\,,
\end{equation}
where $L_{\rm sd}$ is the spindown luminosity of the magnetar, $R$ is the radial distance of the nebula-ejecta boundary from the center of the magnetar, $t_{\rm diff}^{\rm neb}$ and $t_{\rm diff}^{\rm ej}$ are the photon diffusion timescales in the nebula and in the ejecta, respectively, $\mathcal{A}$ is the fraction of non-thermal photons that escape from the system, $\varepsilon_B$ is a parameter that determines the magnetic field strength, and $Q_{\rm heat}^{\rm rp}$ accounts for the rate of heating due to decay of r-process nuclei in the ejecta.

The evolution of the non-thermal, thermal, and magnetic energies depend on the spin-down luminosity of the magnetar ($L_{\rm sd}$), the adiabatic expansion, and the energy transfer between the nebula and ejecta by photons, and the radiative energy losses. We describe these processes in what follows.
We evaluate the physical quantities in the rest frame of the magnetar and use $X'$ for the quantities in the fluid rest frame unless otherwise noted. Although our formulation ignores relativistic effects in some aspects, the expanding nebula and ejecta is at most in the mildly relativistic regime as seen below. Thus, the relativistic correction does not severely affect our results.

\begin{figure*}
\begin{center}
\includegraphics[width=0.48\textwidth]{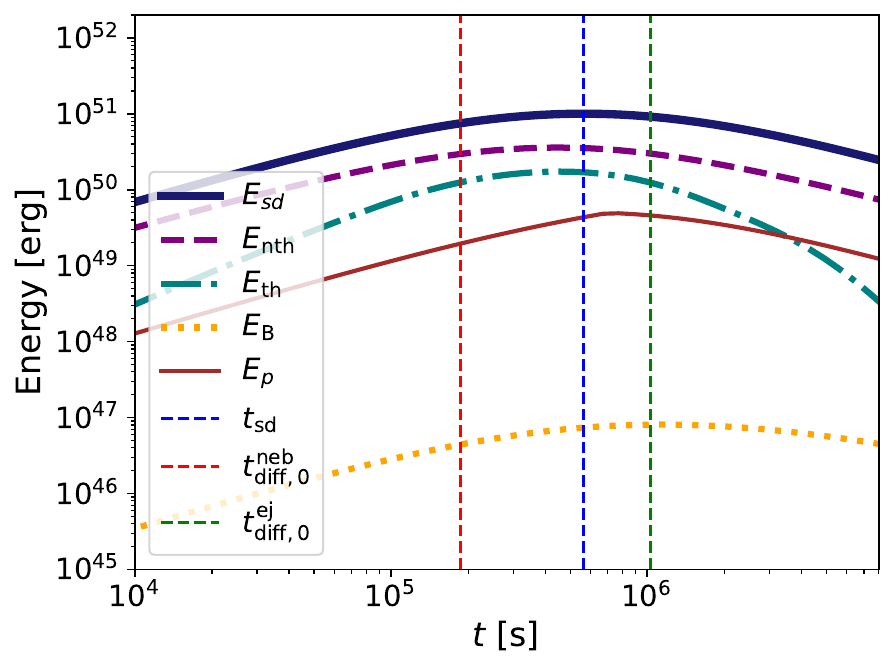}
\includegraphics[width=0.48\textwidth]{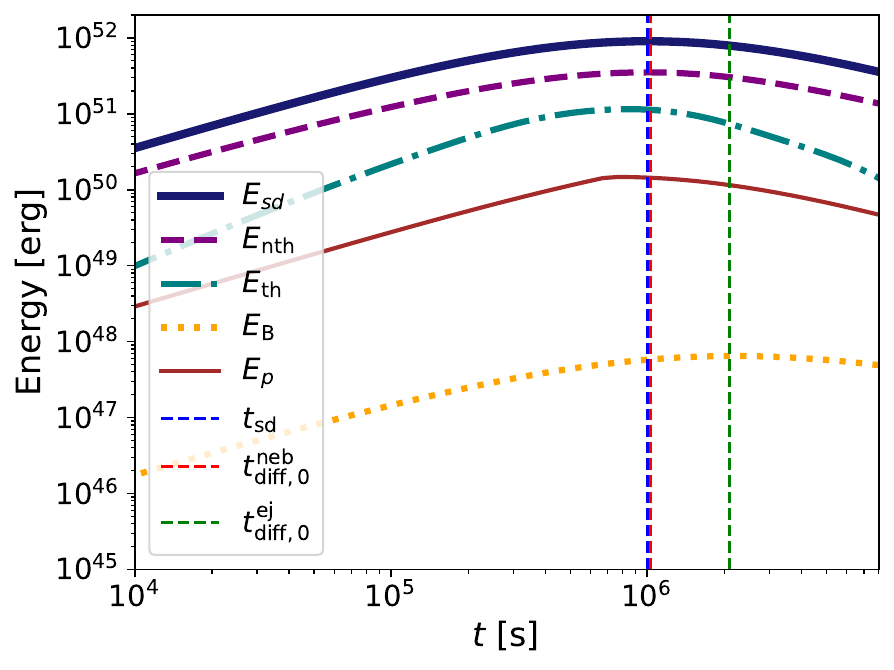}
\caption{\label{fig:energies}Time evolution of the total spindown (solid thick dark blue), non-thermal (dashed purple), thermal (dot-dashed sea-green), magnetic (dotted dark yellow), and CR proton (thin solid brown) energies, along with the initial spin period ($t_{\rm sd}$) (dashed blue), $t_{\rm diff,0}^{\rm neb}$ (dashed red), and $t_{\rm diff,0}^{\rm ej}$ (dashed green) timescales. The \emph{left} panel shows the \emph{fiducial} case, while the \emph{right} panel shows the \emph{optimistic} scenario. Refer to Table~\ref{tab:params} for details on the choice of parameters for both scenarios.
}
\end{center}
\end{figure*}
The first term in Equations (\ref{eq:nth}) and (\ref{eq:eb}) is the magnetar spin-down power $L_{\rm sd}$. The total rotational energy of the magnetar is $E_{\rm rot} = (1/2) I \Omega_i^2$, where $I$ is the moment of inertia which we assume to be $I = (2/5) M_* R_*^2$ and $\Omega_i$ is the initial angular velocity. For an assumed neutron star mass $M_* = 2.3 M_\odot$ and radius $R_* = 10$ km, this implies $I = 1.83 \times 10^{45}\ {\rm g\ 
cm}^{2}$. Unless otherwise stated, the initial spin-period is assumed to be $P_i = 0.003$ s, such that $E_{\rm rot} \sim 4 \times 10^{51}$ erg. This acts as the main energy source.

The newly born magnetar will spin down via both the magnetic dipole torque and the GW quadrupole radiation as a result of deformities of the magnetar. Thus, the spin evolution of the magnetar is given by~\citep{1983bhwd.book.....S}, $\dot{E}_{\rm rot} = - I \dot{\Omega} = L_{\rm sd} + L_{\rm GW}$, where $\Omega$ is the angular velocity, $L_{\rm sd}$ and $L_{\rm GW}$ are the energy injection rates into the nebula and GW radiation respectively. The rotation energy of the magnetar is injected into the nebula with the energy injection rate of~\citep{OstrikerGunn1969}
\begin{align}
\label{eq:lsd}
L_{\rm sd} = \alpha \frac{\mu^2 \Omega^4}{c^3} &= 7.13 \times 10^{45}\ {\rm erg\ s}^{-1} \left( \frac{B_d}{10^{14}\ \rm G} \right)^2 \nonumber\\ 
&\left( \frac{P_i}{0.003\ \rm s} \right)^{-4} \left( 1 + \frac{t}{t_{\rm sd}} \right)^{-2}
\end{align}
where $\mu = B_{d} R_*^3$ denotes the magnetic dipole moment, 
$B_d$ is the surface equatorial dipole field, the angular velocity $\Omega = (2 \pi)/ P$, the spin period $P=P_i \big( 1+t/t_{\rm sd} \big)^{1/2}$, 
and $\alpha$ is a parameter to account for the inclination angle between the rotation and magnetic axes. In general, for an aligned case $\alpha = 1$. The above expression for the spin-down luminosity assumes force-free electrodynamics~\citep{Gruzinov:1999aza,Blandford2002}, which only holds if the electromagnetic energy density is greater than the plasma pressure and inertia and when the plasma drift velocity is subluminal, $E^2 \leq B^2$~\citep{Spitkovsky:2006np}. The violation of the latter condition can lead to instabilities and the force-free equations can no longer be used. The spin-down time can be given by the ratio of the spin energy ($E_{\rm rot}$) and the initial spin-down luminosity ($L_{\rm sd} (t=0)$),
\begin{equation}
t_{\rm sd} = 5.63 \times 10^5\ {\rm s} \left( \frac{B_d}{10^{14}\ \rm G} \right)^{-2} \left( \frac{P_i}{0.003\ \rm s} \right)^2\,.
\end{equation}

The rate of energy injection into GW radiation is given by~\citep{OstrikerGunn1969},
\be
L_{\rm GW} = \frac{32}{5} \left( \frac{G}{c^5} \right) I^2 \epsilon^2 \Omega^6\,,
\ee
where $\epsilon$ defined as the ellipticity is a dimensionless quantity quantifying the deformity of the star and $G$ is Newton's gravitational constant. The ellipticity of the magnetar is dependent on the particulars of the deformation mechanism. A purely toroidal magnetic field can produce large ellipticity, $\epsilon \sim 0.016 (B_t/10^{17}\rm G)^2$~\citep{Stella:2005yz,DallOsso:2008kll}. In this work, we consider poloidal magnetic fields. When poloidal magnetic fields serve as the dominant deformation mechanism, the ellipticity can be approximately estimated to be~\citep{Bonazzola:1995rb,Xie:2022igk},
\be
\epsilon = \frac{45}{64 \pi} \frac{B_d^2}{G \rho^2 R_*^2} = 2.8 \times 10^{-8} \left( \frac{B_d}{10^{14}\ \rm G} \right)^2\,,
\ee
where $\mu_0$ is the magnetic permeability of free space, the constant mass density of the star $\rho = M_*/\left((4/3) \pi R_*^3\right)$. This assumes an incompressible magnetized fluid for simplicity. Thus we have,
\be
\label{eq:lgw}
L_{\rm GW} \approx 5 \times 10^{35}\ {\rm erg\ s}^{-1}\ \left( \frac{\epsilon}{10^{-8}} \right)^{2} \left( \frac{P_i}{0.003\ \rm s} \right)^{-6} \left( 1 + \frac{t}{t_{\rm sd}} \right)^{-3}\,.
\ee
As can be noted from Equations~\eqref{eq:lsd} and~\eqref{eq:lgw}, the rotational energy losses due to quadrupolar radiation via GWs is \emph{negligible} as compared to the energy loss in the electromagnetic channel. In fact, even if one argues that the ellipticity is uncertain, the energy losses only become comparable when $\epsilon > 10^{-5}$. Thus, for all purposes we only consider spindown through the electromagnetic channel. Furthermore,~\cite{Arons:2002yj} also consider the possibility of GW emission resulting from r-mode instabilities in the magnetar, given by~\citep{Owen:1998xg}
\be
L_{\rm GW}^{\rm r-mode} = \frac{512}{315} \left( \frac{GI^2 \Omega^6}{c^5} \right) \left( \frac{\Omega R_*}{c} \right)^2 \alpha_{\rm r-mode}^2\,,
\ee
where $\alpha_{\rm r-mode}$ is a dimensionless mode amplitude. For typical values of $\alpha_{\rm r-mode}$ the effect on the CR proton spectrum is above $10^{12}$ GeV, where it might lead to a very hard cutoff. This is because a fraction of the spindown energy is lost in GW losses because of this r-mode instability and hence cannot be used for acceleration of the CR protons to such high energies. Since in this work we do not accelerate CR protons to such high energies (see Section~\ref{subsec:cr_prot} for details), we can neglect the GW emission or losses due to r-mode instabilities of the magnetar.

The magnetic field in the nebular region is determined by dissipation processes, such as shocks and turbulence. This is incorporated in our formalism using $\varepsilon_B$ in Equation~\eqref{eq:eb}, where the nebular magnetic field strength is given by $\varepsilon_B$. We choose $\varepsilon_B \sim 10^{-4}$ motivated by recent results from~\cite{Vurm:2021dgo}. Note that this is lower than what was found in axisymmetric two-dimensional numerical simulations to model the  Crab Nebula observations~\citep{Komissarov:2003tg,DelZanna:2004aq,Bogovalov2005,2010ApJ...715.1248T}, where $\epsilon_B \sim 10^{-2}$. This value has uncertainties associated with it (see \citealt{Komissarov:2012sr} for details). For larger values of $\epsilon_B$, the synchrotron emission can dominate the radiation field in the nebular region, which is inconsistent with present observations (for example SNe 2015bn~\citealt{Nicholl:2018cam} and 2017egm~\citealt{Nicholl:2017mnb}). We thus choose a value that lies between $10^{-6}$ and $10^{-2}$. The turbulent magnetic field component in the nebular region can be treated as a fluid, because its adiabatic index is same as the relativistic fluid. 

The internal energy of the nebula is dominated by the photons in the nebula, and the magnetic field in the nebula provides a sub-dominant contribution. The internal energy of the nebula under these conditions is given by $E_{\rm neb} = 3 p V$, where $p$ and $V$ are nebular pressure and volume respectively. The magnetic field strength ($B_{\rm neb}^\prime$) can be estimated to be 
\begin{equation}
\label{eq:bneb}
B^\prime_{\rm neb} = \left(\frac{ 8 \pi E_B}{ 4 \pi R^2 \Gamma_{\rm ej}^2 \beta_{\rm ej} c t } \right)^{1/2}\,.
\end{equation}

The second term in equations~\eqref{eq:nth},~\eqref{eq:th}, and~\eqref{eq:eb} represents the adiabatic expansion term, which affects the outward expansion of the nebula-ejecta system. The kinetic energy of the ejecta ($E_{\rm kin}$) depends on the thermal and non-thermal emissions and is given by
\begin{equation}
\frac{d}{dt} E_{\rm kin} = \frac{d}{dt} \left[ M_{\rm ej}c^2 (\Gamma_{\rm ej} - 1) \right] = \frac{v}{R} \left( E_{\rm nth}+E_{\rm th}+E_B \right)\,,
\end{equation}
where $M_{\rm ej}$ is the mass of the ejecta, $c$ is the speed of light, $\Gamma_{\rm ej}$ is the Lorentz factor associated with the ejecta, $v = dR/dt$, is the velocity of the ejecta. The mean radius evolves as
\begin{equation}
R(t) = R_{\rm init} + \int_{t_{\rm in}}^{t} d\tilde{t}\ v\,,
\end{equation}
where $R_{\rm init}$ and $t_{\rm in}$ are the initial mean radius and time respectively.

The third term in Equation~\eqref{eq:nth} takes into account the conversion of non-thermal photons to thermal radiation, which acts as a source for the thermal radiation accounted for by the first term in Equation~\eqref{eq:th}. The reprocessing occurs over a timescale which is decided by the time it takes for the photons to traverse from the nebula to the inner boundary of the ejecta. This sets the diffusion timescale of the photons ($t_{\rm diff}^{\rm neb}$) given by
\begin{align}
t_{\rm diff}^{\rm neb} &\approx \frac{R}{c} (1+\tau_{\rm diff}^{\rm neb}) \nonumber\\
\tau_{\rm diff}^{\rm neb} &= \frac{n_{\pm}' \sigma_T R}{\Gamma_{\rm ej}} = \left( \frac{4 Y L_{\rm sd} \sigma_T}{\pi m_e c^3 R \Gamma_{\rm ej}^2} \right)^{1/2}
\,,
\end{align}
where $\tau_{\rm diff}^{\rm neb}$ is the Thomson optical depth in the nebular region, $\sigma_T$ is the Thomson optical depth, $m_e$ is the mass of an electron, and $n_{\pm}'$ is the number density for $e^{\pm}$ pairs in the nebula rest frame, and $Y$ is the pair multiplicity. The above expression for the optical depth is calculated assuming a balance between the pair creation and annihilation rates. The pair multiplicity factor $Y$ is chosen to be $0.1$ assuming a saturated state~\citep{1987MNRAS.227..403S}.

The ejecta and nebula expand on a timescale, $t_{\rm ej}^{\rm exp} = R/\left( \beta c \right)$, which gives the timescale of adiabatic loss. Initially, $t_{\rm diff}^{\rm neb} > t_{\rm ej}^{\rm exp}$ is satisfied, and the adiabatic loss is significant. As the ejecta expands, $t_{\rm diff}^{\rm neb}$ decreases and $t_{\rm ej}^{\rm exp}$ increases, and eventually, photons start to diffuse out efficiently. We define $t_{\rm diff,0}^{\rm neb}$ at the time when $t_{\rm diff}^{\rm neb} = t_{\rm ej}^{\rm exp}$, or equivalently, $\tau_{\rm diff}^{\rm neb} = 1/\beta$ is satisfied. 

Similar to the photons diffusing through the nebula that get reprocessed into thermal radiation, the thermal photons diffuse (and escape) through the ejecta over a timescale ($t_{\rm diff}^{\rm ej}$) given by,
\begin{align}
\label{eq:ejjdiff}
t_{\rm diff}^{\rm ej} &\approx \frac{R}{c} (1+\tau_{\rm diff}^{\rm ej}) \nonumber\\
\tau_{\rm diff}^{\rm ej} &= \frac{\rho_{\rm ej}^\prime R \kappa_{\rm ej}}{\Gamma_{\rm ej}} = \frac{3 M_{\rm ej} \kappa_{\rm ej}}{4 \pi R^2 \Gamma_{\rm ej}^2}\,,
\end{align}
where $\rho_{\rm ej}^\prime = M_{\rm ej}/\left( (4/3) \pi R^3 \Gamma_{\rm ej} \right)$ is the number density of the ejecta in the fluid-rest frame, $\tau_{\rm diff}^{\rm ej}$ is the optical depth for the ejecta, $\kappa_{\rm ej}$ is the opacity in the ejecta. We choose $\kappa_{\rm ej} = 10\ \rm cm^2 g^{-1}$, considering that the ejecta is filled with a r-process elements~\citep{Tanaka:2019iqp,Fujibayashi:2020dvr}. Similar to the nebular case, the photons can diffuse out from the ejecta when $t_{\rm diff}^{\rm neb} \leq t_{\rm ej}^{\rm exp}$, and we define $t_{\rm diff,0}^{\rm ej}$ at the time when the photons start to diffuse out efficiently. Note that we assume that the nebula and the ejecta have the same values of Lorentz factor and physical size, $\Gamma_{\rm ej}$, and $R$.

The nebula produces copious amounts of UV and X-ray photons as a non-thermal component. When these photons reach the ejecta (in this case the mean radius $R$), a fraction $\mathcal{A}$ of them escape and the rest ($1-\mathcal{A}$) are reprocessed into thermal radiation. The quantity $\mathcal{A}$ can be computed using the ratio of the photon luminosity diffusing out from the nebula post attenuation to the photon luminosity escaping from the ejecta post attenuation. We compute $\mathcal{A}$ consistently as a function of time\footnote{In~\cite{Fang:2017tla}, $\mathcal{A}$ was assumed to be $0$.}. 
We take into account the free-free absorption in radio band, a Planck-mean bound-bound opacity in infrared/optical/ultraviolet band, a wave-length dependent bound-free and Compton scattering opacity in the X-ray to soft gamma-ray band, and Bethe-Heitler opacity in high-energy gamma-ray band, details of which will be discussed in a companion paper~\citep{Mukhopadhyay:2025tvz}.

The last term in Equation~\eqref{eq:th} accounts for the heating of the ejecta and the nebula due to the decay of r-process elements. The energy released as a result of these decays can power kilonovae emissions. We take this into account to consistently evaluate the thermal energy in the nebula. We follow the prescription detailed in~\cite{Hotokezaka:2015cma} to make some rough estimates and quantify this. The heating rate can be defined as
\begin{align}
Q^{\rm heat}_{\rm rp} &=  f_{\rm therm} f_{\rm diff} \dot{E}_{\rm rp} M_{\rm ej}\, \nonumber\\
\dot{E}_{\rm rp} &= \dot{E}_{0,\rm rp} \left( \frac{t}{t_{0,\rm rp}} \right)^{-1.3}\,,
\end{align}
where $f_{\rm therm}$ is defined as the fraction of photons that have thermalized and hence can contribute to the thermal energy in the nebula. The evaluation of $f_{\rm therm}$ is non-trivial. For simplicity, we choose\footnote{This choice is based on results from~\cite{Hotokezaka:2015cma} (see Figure 3 there).}, $f_{\rm therm}$ = 0.7 for $t < t_{\rm tr,\gamma}$ and 0.2 for $t>t_{\rm tr,\gamma}$, where the time when ejecta becomes transparent to gamma-rays is defined as $t_{\rm tr,\gamma} \approx \big( \kappa_\gamma M_{\rm ej}/(4 \pi v^2) \big)^{1/2} \approx 0.4\ {\rm day} \big( \kappa_\gamma/0.05\ {\rm cm}^2 {\rm g}^{-1} \big)^{1/2}  \big( M_{\rm ej}/0.01 M_\odot \big)^{1/2} \big( \beta_{\rm ej}/0.3 \big)^{-1}$. The opacity of gamma-rays is given by $\kappa_\gamma$ and we choose it to be $0.05\ {\rm cm}^2 {\rm g}^{-1}$. Only a fraction of the thermalized photons can diffuse in to the nebula. We quantify this fraction $f_{\rm diff} = t_{\rm dyn}/t_{\rm diff}^{\rm ej}$, which is essentially the fraction of photons that diffuse from the ejecta to the nebula, $t_{\rm dyn} = R/\big( \Gamma_{\rm ej} v \big)$ is the dynamical timescale associated with the ejecta, $t_{\rm diff}$ is defined in Equation~\eqref{eq:ejjdiff}. At early times, the ejecta is optically thick and hence $f_{\rm diff}$ is expected to be low. At late times the ejecta becomes optically thin and $f_{\rm diff}$ increases. We choose the normalizations $\dot{E}_{0,\rm rp} = 1.6 \times 10^{11}\ {\rm erg}\ {\rm s}^{-1} {\rm g}^{-1}$ and $t_{0,\rm rp} = 10^4$ s. Numerical simulations reveal that the energy generation as a result of the decay processes roughly behave as a power law with time $\sim t^{-1.3}$~\citep{Metzger:2010sy}.

\begin{table}
\centering
\caption{Table of parameters}
\label{tab:params}
\begin{center}
Shared parameters
\end{center}
\begin{minipage}{0.48\textwidth}
\centering
\begin{tabular}{cccccccc}
\hline
\hline
$M_*$ & $R_*$ & $\alpha$ & $\epsilon_B$ & $Y$ & $\kappa_{\rm ej}$ & $\eta_{\rm gap}$ & $\eta_{\rm acc}$\\
\hline
$2.3\ M_\odot$  & $10$ km & 1 & $10^{-4}$ &  0.1 & 10 cm$^{-2}$ g$^{-1}$ & 0.1 & 1 \\
\hline
\end{tabular}
\end{minipage}
\begin{center}
Parameter sets for the fiducial and optimistic models
\end{center}
\begin{minipage}{0.30\textwidth}
\centering
\begin{tabular}{cccc}
\hline
\hline
Parameter & Fiducial & Optimistic \\
\hline
$P_i$ & $3$ ms & $1$ ms \\
$B_d$ & $10^{14}$ G & $2.5 \times 10^{13}$ G \\
$M_{\rm ej}$ & $0.03\ {\rm M}_\odot$ & $0.1\ {\rm M}_\odot$ \\
$v_0$ & $0.2$ c & $0.1$ c \\
$f_{\rm mag}$ & $ 0.1 $ & $ 0.3 $ \\
\hline 
\end{tabular}
\end{minipage}
\end{table}
At this stage, we have discussed all the terms that appear in equations~\eqref{eq:nth},~\eqref{eq:th}, and~\eqref{eq:eb}. We examine two sets of parameters: a moderate set (\emph{fiducial}) and an \emph{optimistic}\footnote{Here \emph{optimistic} implies the model parameters suited for an efficient neutrino-production scenario.} set, the values of which are provided in Table~\ref{tab:params}. 
In Figure~\ref{fig:energies}, we show the total spindown (thick solid dark blue), non-thermal (dashed purple), thermal (dot-dashed sea-green), magnetic (dotted dark yellow), and cosmic ray (CR) proton (thin solid brown) energies along with the spindown timescale (dashed blue), the characteristic timescales for nebular (dashed red) and ejecta (dashed green) diffusion. 

The characteristic nebular diffusion time is less than the spindown time for the fiducial case, whereas the two are very similar with the characteristic nebular diffusion time being only slightly larger than the spindown time for the optimistic case. The total spin-down energy\footnote{Note that we choose to define $E_{\rm sd}$ in this way instead of integrating it over time because this helps in visualizing the evolution of $E_{\rm sd}$ over particular time intervals. Computing $E_{\rm sd}$ as a time-integral would lead to the results being cummulative which does not highlight the time structure of spindown energy with respect to the spindown timescale ($t_{\rm sd}$). Besides, this is also a fairly good approximation where the time evolution is governed by a power-law which is the case here. We have checked that the agreement between the two ways of doing this is very good (upto a factor of $\sim 2$).}, $E_{\rm sd} = L_{\rm sd} t$, increases and peaks at the initial spin-down time ($t_{\rm sd}$). Following this, the spin-down energy goes down. Owing to the smaller initial spin period, the optimistic scenario has a larger spindown energy ($E_{\rm sd}$) by roughly one order of magnitude. 

The non-thermal energy closely follows the behavior of the spindown energy and peaks around $t_{\rm sd}$. This is expected since the spin energy of the magnetar is the only energy source for the non-thermal photons in the nebula. There are some subtle differences between the peak of the spin-down energy and the non-thermal energy owing to $t>t_{\rm diff}^{\rm neb}$. The non-thermal energy dominates over the thermal and the magnetic energies throughout the time evolution.

The thermal energy peaks slightly earlier than the non-thermal energy owing to the reprocessing of non-thermal photons to thermal ones by the nebula-ejecta boundary (the first term in Equation~\ref{eq:th}). More importantly, the thermal energy rapidly decreases for $t>t_{\rm diff,0}^{\rm ej}$ when the ejecta becomes transparent to thermal photons (the third term in Equation~\ref{eq:th}). The magnetic energy also follows the time evolution of the total spindown energy since $dE_B/dt \propto L_{\rm sd}$ and peaks around $t_{\rm sd}$. The peak magnetic field energy is $\sim 8 \times 10^{46}$ erg and $8 \times 10^{47}$ erg for the fiducial and the optimistic cases respectively.

For our choice of fiducial and optimistic parameters, the contribution from the heating resulting from the decay of r-process elements is insignificant and can be ignored. This is because at initial times although $f_{\rm therm}$ is large, $f_{\rm diff}$ is very low since the ejecta is optically thick. At late times, first of all $Q_{\rm rp}^{\rm heat}$ falls off as $\sim t^{-1.3}$, secondly, $f_{\rm therm}$ is $\sim 0.2$. This results in insignificant contribution of this term throughout the time evolution of the system. It is possible that for larger $B_d$ ($>10^{15}$ G) and higher ejecta masses, $Q_{\rm rp}^{\rm heat}$ might dominate during late times.

\begin{figure}
\includegraphics[width=0.48\textwidth]{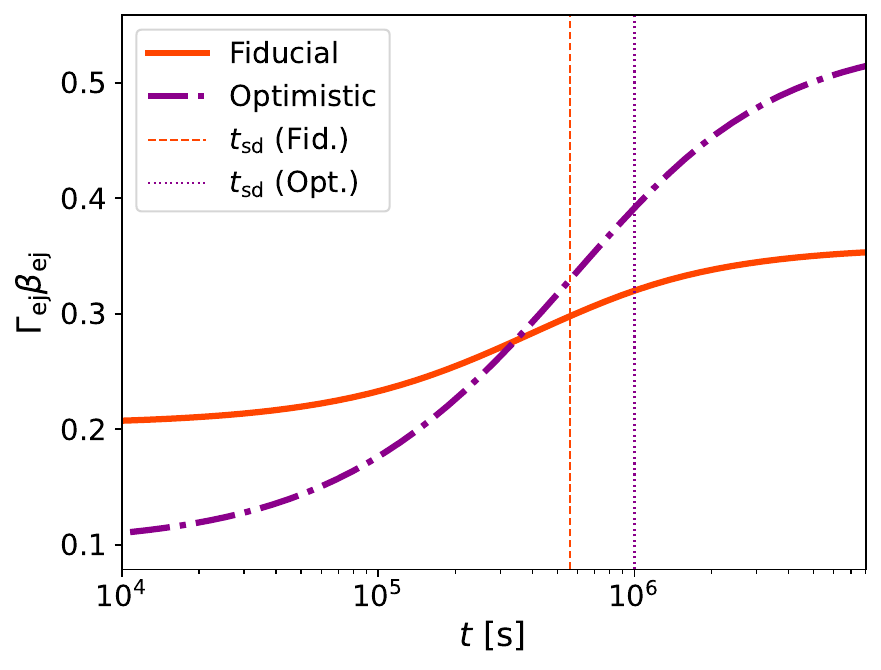}
\caption{\label{fig:gammabeta} Time evolution of $\Gamma_{\rm ej} \beta_{\rm ej}$ for the fiducial case and the optimistic scenario. Refer to Table~\ref{tab:params} for details on the choice of parameters for both scenarios.
}
\end{figure}

In Figure~\ref{fig:gammabeta} we show the time evolution of $\Gamma_{\rm ej}\beta_{\rm ej}$ for the fiducial (optimistic) case in solid red-orange (dot-dashed purple) line. This gives us an insight regarding the evolution of the nebula-ejecta boundary. At initial times, the value of $\Gamma_{\rm ej}\beta_{\rm ej}$ corresponds to the initial velocity of $v_0 = 0.2c~(0.1c)$ for the fiducial (optimistic) case. The boundary and hence the ejecta undergoes an accelerated expansion close to the spindown time. At later phases, $\Gamma_{\rm ej}\beta_{\rm ej}$ saturates and expands with a constant velocity with a mildly relativistic speed. Even for the optimistic scenario which has $E_{\rm sd}$, the relativistic effect is subtle as $\Gamma_{\rm ej}\beta_{\rm ej}\sim 0.6$.
\section{Neutrino production}
\label{sec:nu_prod}

High-energy neutrino production in astrophysical systems is generally sourced by hadronuclear ($pp$) or photomesonic ($p\gamma$) processes. This requires the acceleration of ions (protons or nuclei) to sufficiently high energies. The surface of the magnetar acts as the source of the ions or cosmic-ray (CR) protons. We discussed in Section~\ref{sec:model} that there are two proton acceleration sites in our model - the potential gap at the polar cap and the termination shock (TS) of the wind. In the former scenario, the acceleration of ions is achieved when the ions travel along the magnetic field lines in the pulsar magnetosphere~\citep{2013MNRAS.429...20T}. The latter scenario involves acceleration of ions due to the shock or reconnection~\citep{Sironi:2011zf,Sironi:2014jfa}. We only consider CR protons in this work and hence in all cases $Z=1$ (see Section~\ref{sec:disc} for a discussion on the effects of including heavier nuclei).

\subsection{Acceleration of CR protons}
\label{subsec:cr_prot}
In this sub-section we discuss the acceleration of the CR protons. We begin by discussing the injection spectrum associated with CR protons. We explore the cut-off energy for the acceleration of the protons in the polar cap region. We also discuss the electron and photon spectra relevant for studying the resulting neutrino spectra along with the various relevant timescales in this sub-section.
\subsubsection{CR proton spectrum}
After the BNS merger, baryons are ejected as a kilonova ejecta, and the remnant magnetar sources pulsar winds. In this phase, the dense plasma around the neutron star is blown away, forming a magnetosphere around the remnant magnetar. The number density of charges in the remnant magnetar is estimated to be
$n_{\rm GJ} = - \bm \Omega\cdot\bm B/\big(2 \pi Z e c\big)$~\citep{Goldreich:1969sb}, where $Z$ is the atomic number, $e$ is the charge of an electron. We use this to define the CR (proton) production rate as follows
\begin{equation}
\dot{N}_p = n_{\rm GJ}2 A_{\rm pc} c = \frac{4 \pi^2}{Z e} \frac{R_*^3}{c} \frac{B_0}{P^2}\,,
\end{equation}
where $A_{\rm pc}= \pi R_*^2 \big( R_*/R_{\rm lc} \big)$ is the size of the polar cap, and $R_{\rm lc} = c/\Omega$ is the light cylinder radius. Assuming an aligned case such that, $\bm \Omega\cdot \bm B = \Omega B$ gives the final expression.
The protons can then be accelerated across multiple sites. In particular, we consider the polar cap region and the termination shock (TS) (see Section~\ref{subsec:timescales}) region as potential acceleration sites for the CR protons.

The particle accelerations at the polar-cap region can be modeled using a Dirac-delta function spectrum~\citep{Arons:2002yj} whereas the termination shock acceleration should result in a power-law CR spectrum \citep[e.g.,][]{Blandford:1987pw,Guo:2020fni}. The polar cap region can efficiently accelerate all CR protons in between an energy window given as a function of the cut-off energy in the polar cap region $\varepsilon_p^{\prime \rm cutoff, pc}$. The resulting CR injection spectra can then be approximately given by
\begin{equation}
\label{eq:crspec_inj}
\frac{d\dot{N}_{p,\rm inj}}{d \varepsilon_p^\prime } = \dot{N}_p^{\rm norm} Q_p^{\rm inj}(\varepsilon_p^\prime ) = \dot{N}_p\delta\big(\varepsilon_p^\prime -  \varepsilon_p^{\prime \rm cutoff, pc} \big)\,,
\end{equation}
where the normalization $\dot{N}_p^{\rm norm}$ is given by $\dot{N}_p^{\rm norm} = \dot{N}_{p}/\big( \int_{\varepsilon_p^{\rm min}}^\infty d \varepsilon_p^\prime  Q_p^{\rm inj}(\varepsilon_p^\prime ) \big)$ and $\varepsilon_p^{\prime \rm cutoff,pc}$ is the cut-off energy of the CR protons in the polar cap region (see Section~\ref{subsec:polarcap_en}). Note that since we have $\delta$-function injection spectrum, it is by default normalized to 1, that is, $\int_{\varepsilon_p^{\rm min}}^\infty d \varepsilon_p^\prime  Q_p^{\rm inj}(\varepsilon_p^\prime ) = 1$ and thus, $\dot{N}_p^{\rm norm} = \dot{N}_{p}$. The minimum energy of the CR protons\footnote{Since we have a $\delta-$function CR proton injection spectra, this choice does not affect our results in any way.} is chosen to be, $\varepsilon_p^{\rm min}=\Gamma_{ej}m_pc^2 \sim 1$  GeV. The dot denotes the time-derivative.

The steady state CR proton spectrum is computed by solving the transport Equation~\citep[e.g.,][]{Kimura:2019yjo}
\begin{equation}
\frac{d}{d \varepsilon_p^\prime}\left( -\frac{\varepsilon_p^\prime}{t_{\rm cool}^\prime} \frac{d N_p}{d\varepsilon_p^\prime}\right) = \frac{d \dot{N}_{p,\rm inj}}{d\varepsilon_p^\prime} - \frac{1}{t_{\rm esc}^\prime} \frac{d N_p}{d \varepsilon_p^\prime}\,,
\end{equation}
where the injection spectrum $d\dot{N}_{p,\rm inj}/d\varepsilon_p^\prime$ is defined in Equation~\eqref{eq:crspec_inj}, $t_{\rm esc}^\prime$ is the escape timescale. The transport equation has an analytic solution given by~\citep{DermerMenon}
\begin{align}
\frac{d N_p}{d\varepsilon_p^\prime} = \frac{t_{\rm cool}^\prime}{\varepsilon_p^\prime} \int_{\varepsilon_p^\prime}^{\infty} &d \tilde{\varepsilon}_p \dot{N}_{p,\rm inj} (\tilde{\varepsilon}_p)\ {\rm exp} \bigg( -\mathcal{G}(\varepsilon_p^\prime,\tilde{\varepsilon}_p) \bigg)\, \nonumber\\
\mathcal{G}(\varepsilon_1,\varepsilon_2) & = \int_{\varepsilon_1}^{\varepsilon_2} \frac{t_{\rm cool}^\prime}{t_{\rm esc}^\prime}\frac{d \tilde{\varepsilon}_p}{\tilde{\varepsilon}_p}\,,
\end{align}
where $\tilde{\varepsilon}_p$ is the integration variable and the integration is performed numerically.

The total energy in CR protons ($E_p$) is defined as\footnote{We define $E_p$ in this way instead of doing a time integral (similar to $E_{\rm sd}$) because of the same reasons.}
\begin{equation}
E_p = \dot{N}_{p,\rm inj} \varepsilon_p^{\prime \rm cutoff,pc} t\,.
\end{equation}

The total energy in CR protons ($E_p$) is shown as a solid thin brown line in Figure~\ref{fig:energies} for the fiducial (left) and optimistic case (right). The shifted peak of $E_p$ is because of a combination of factors: $\dot{N}_p$ decreases for $t>t_{\rm sd}$, but the time-integrated value $N_p = \int dt \dot{N}_p$ saturates for $t>t_{\rm sd}$. However, the cut-off energy for the protons $\varepsilon_p^{^\prime \rm cutoff,pc}$ has a peak later in time, that is, for $t>t_{\rm sd}$. We also see that the optimistic scenario has an order of magnitude higher CR proton energy than the fiducial case. This is a result of the smaller initial spin period ($P_i$). As a result of the longer spin-down timescale in the optimistic scenario as compared to the fiducial scenario, $E_p$ peaks at a later time for the former in comparison to the latter. This has an impact in the neutrino fluences which is seen to peak later for the optimistic scenario and last longer as will be discussed in Section~\ref{sec:res}.
%
\subsubsection{Maximum energy from polar cap acceleration}
\label{subsec:polarcap_en}
\begin{figure*}
\begin{center}
\includegraphics[width=0.48\textwidth]{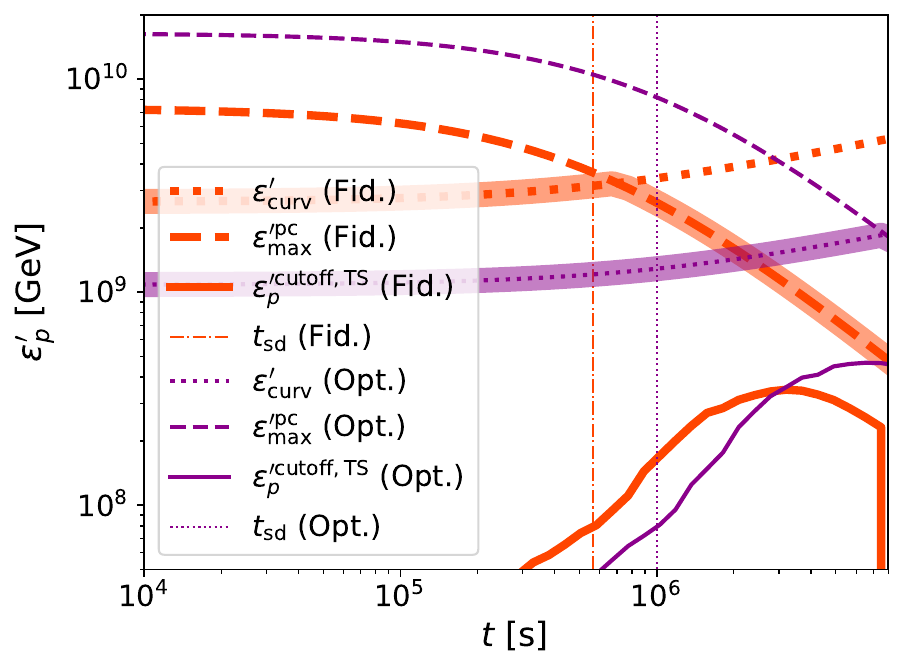}
\includegraphics[width=0.48\textwidth]{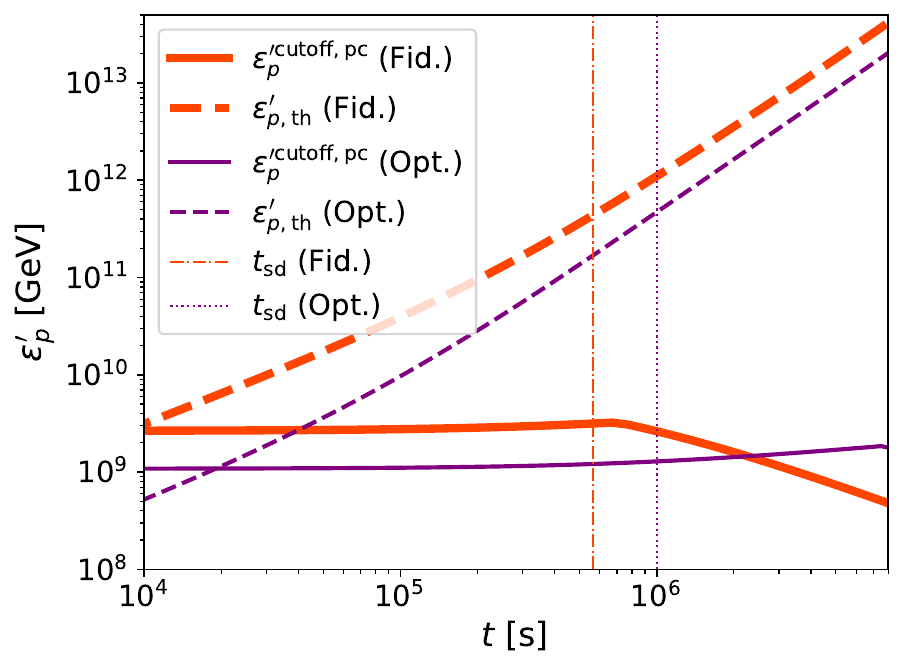}
\caption{\label{fig:picooling} \emph{Left: }Time evolution of the relevant CR proton energies in different regions: the maximum energy that the protons can be accelerated to in the polar cap (see Equation~\ref{eq:epcmax}) and TS (see Section~\ref{subsec:timescales}) regions are denoted by $\varepsilon_{\rm max}^{\prime \rm pc}$ and $\varepsilon_p^{\prime \rm cutoff, TS}$ respectively, the proton energy limited by curvature losses (see Equation~\ref{eq:ecurv}) in the polar cap region is denoted by $\varepsilon^\prime_{\rm curv}$, for the fiducial (Fid.) and optimistic (Opt.) scenarios. The cut-off energy in the polar cap region $\varepsilon_p^{\prime \rm cutoff,pc}$ (see Equation~\ref{eq:epcutoffpc}) appearing in the injection spectrum (see Equation~\ref{eq:crspec_inj}) is shown as a shaded band. \emph{Right: }Time evolution of the cut-off energy of the protons in the polar cap region $\varepsilon_p^{\prime \rm cutoff, pc}$ (see Equation~\ref{eq:epcutoffpc}) and the threshold proton energy associated with pion decay $\varepsilon_{p,\rm th}^\prime = 0.2 \varepsilon_{\pi,\rm th}^\prime$ (see Equation~\ref{eq:epithreshold}), for the fiducial (Fid.) and the optimistic (Opt.) scenarios. Refer to Table~\ref{tab:params} for details on the choice of parameters for both scenarios. 
}
\end{center}
\end{figure*}
The relativistic outflowing plasma in the presence of the rapidly rotating magnetar along with the strong dipolar field leads to a potential gap, $\Phi_{\rm mag} = \Omega^2 \mu/c^2$. Charged particles ($Ze$) can be accelerated in such potential differences up to energies~\citep{Arons:2002yj,Blasi:2000xm,Fang:2013vla}
\begin{equation}
\label{eq:epcmax}
\varepsilon^{\prime \rm pc}_{\rm max}= 4 \eta_{\rm gap} (Ze) B_d \left( \frac{\pi R_*}{c P} \right)^2 R_*\,,
\end{equation}
where we assume that the charged particle experiences a fraction $\eta_{\rm gap}$ of the potential gap. We choose $\eta_{\rm gap} = 0.1$. This kind of potential gap acceleration of charged particles happens only when the density of plasma is really low, such as magnetosphere of compact objects. There is an equivalent way to arrive at Equation~\eqref{eq:epcmax} using the Hillas criterion which limits the maximum energy to which charges can be accelerated given the physical size of the system. From the Hillas criterion, we have, $\varepsilon^{\prime \rm pc}_{\rm max} = (Ze) B R_L$, where the physical size of the system is given by the Larmor radius $R_L$ and is assumed to be the light cylinder radius $R_{\rm lc} (= c/\Omega)$ and the magnetic field at the light cylinder radius, $B = B_d \left( R_*/R_{\rm lc} \right)^3$ (since dipolar magnetic field $\propto 1/r^3$). Putting it altogether, and assuming the screening effect to be $\eta_{\rm gap}$ we arrive at the same equation as above. We show the time-evolution of $\varepsilon^{\prime \rm pc}_{\rm max}$ (dashed) in the left panel of Figure~\ref{fig:picooling} for the fiducial (thick red-orange) and the optimistic (thin purple) scenarios. The maximum polar cap energy is constant at initial times and starts decreasing around the spindown time. This is because $\varepsilon^{\prime \rm pc}_{\rm max} \propto 1/P^2 \propto (1+t/t_{\rm sd})^{-1}$, which implies, at early times it is approximately constant and at later times it falls of as $\sim 1/t$.

The proton energy limited by curvature energy losses is given by $\varepsilon_{\rm curv}^\prime$. The protons lose energy as they move along the curved trajectory due to the dipolar magnetic field lines. The total curvature emission power emitted by the protons can be given by $P_{\rm curv}^\prime = 2 (Ze)^2 \gamma_p^4 c/ \big( 3 R_{\rm curv}^2 \big)$, where $\gamma_p$ is the Lorentz factor of the protons and $R_{\rm curv}$ is the radius of curvature, which we assume to be $R_{\rm lc}$. The timescales associated with curvature loss $t_{\rm curv}^\prime = \gamma_p m_p c^2/P_{\rm curv}^\prime$. The acceleration timescale at the polar cap region can be written as $t_{\rm acc}^{\prime \rm pc} = \varepsilon^\prime/\big( Ze B c \big)$ (assuming $dE/dt \sim Ze B c$). Equating $t^\prime _{\rm curv}$ to $t_{\rm acc}^{\prime \rm pc}$ gives $\gamma_p = \big( 3 B_d R_{\rm curv}^2/(2 Ze) \big)^{1/4}$. Thus the proton energy limited by curvature loss can be expressed as
\begin{equation}
\label{eq:ecurv}
\varepsilon_{\rm curv}^\prime = \gamma_p m_p c^2 = \left[ \frac{3 m_p^4 c^8 B_d R_{\rm curv}^2}{2 Ze}\right]^{1/4}\,.
\end{equation}
The time-evolution of $\varepsilon_{\rm curv}$ (dotted) of the CR protons is shown in Figure~\ref{fig:picooling} (left panel) for the fiducial (thick red-orange) and the optimistic (thin purple) scenarios. It changes by a factor of a few between the initial and final times. This is because $\varepsilon_{\rm curv}^\prime \propto R_{\rm curv}^{1/2} \propto R_{\rm lc}^{1/2} \propto \Omega^{-1/2} \propto P^{1/2} \propto (1+t/t_{\rm sd})^{1/4}$. This implies that at $t << t_{\rm sd}$, $\varepsilon_{\rm curv}^\prime$ is roughly constant and for $t>>t_{\rm sd}$ it increases with $\sim t^{1/4}$.

Finally, the cut-off energy ($\varepsilon_p^{\prime \rm cutoff, pc}$) for the protons in this region is decided by the minimum of $\varepsilon^{\prime  \rm pc}_{\rm max}$ and $\varepsilon_{\rm curv}^\prime$, that is 
\begin{equation}
\label{eq:epcutoffpc}
\varepsilon_p^{\prime \rm cutoff, pc} = {\rm  min} \left[\varepsilon^{\prime  \rm pc}_{\rm max}, \varepsilon_{\rm curv}^\prime  \right]\,.
\end{equation}
%
\subsection{Electron and photon spectra}
\label{subsec:elec_phot_spectra}
In this subsection we discuss the electron and the photon spectra used for the calculation of the neutrino fluences. The spectral energy distribution (SED) fittings of Galactic pulsar-wind nebula (PWNe) reveal that broken power-law injection spectra of non-thermal electrons can explain multiwavelength photon spectra \citep[e.g.][]{2010ApJ...715.1248T}. We use a similar broken power-law injection spectrum for non-thermal electrons:
\begin{equation}
\frac{dN}{d\gamma_e} \sim \ 
\begin{cases}
\gamma_e^{-1.5},\ \gamma_e \leq \gamma_{e,\rm br}\\
\gamma_e^{-2.5},\ \gamma_e > \gamma_{e,\rm br}\,,
\end{cases}
\end{equation}
where $\gamma_{e,\rm br}$ is the break Lorentz factor for the electrons. We choose a harder spectra for $\gamma_e<\gamma_{e,\rm br}$ and a softer spectra for $\gamma_e$ above $\gamma_{e,\rm br}$. For Galactic PWNe, $\gamma_{e,\rm br} \sim 10^5-10^6$ to explain the observed multi-wavelength spectra, which might correspond to the bulk Lorentz factor of the pulsar wind. However, ~\citet{Vurm:2021dgo} showed that the pair injection even at the upstream of the termination shock will cause a gradual decrease of wind velocity, which leads to a lower break Lorentz factor. In addition, recent ALMA observations for superluminous supernovae (SLSNe)\footnote{The magnetar model discussed here, or in general such pulsar-wind nebula models can inject energy into the optically thick ejecta leading to a SLSNe~\citep{Kasen:2009tg}.} resulted in non detection of mm-wave radio signals, which indicates that the values of $\gamma_{e,\rm br}$ and $\epsilon_B$ are lower than those for Galactic PWNe~\citep{2021MNRAS.508...44M}. Hence, we choose $\gamma_{e,\rm br} = 10^3$.

The photon spectra is denoted by $d N_\gamma^\prime/d \varepsilon_\gamma^\prime$. We consider non-thermal and thermal photon fields, which are supplied by the nebula and the ejecta, respectively. The energy density of these photon fields are obtained by solving equations~\eqref{eq:nth} and~\eqref{eq:th}. To obtain the energy distribution of the non-thermal photon field, we solve the transport equation of electron-positron pairs in the nebula and calculate electromagnetic cascades via synchrotron and inverse Compton emissions as well as the subsequent Breit-Wheeler process ($\gamma+\gamma\rightarrow e^++e^-$). The distribution of the thermal photon field is given by estimating the temperature of the ejecta. Details of the photon fields will be discussed in the companion paper~\cite{Mukhopadhyay:2025tvz}.
\subsection{Timescales}
\label{subsec:timescales}
\begin{figure*}
\begin{center}
\includegraphics[width=0.32\linewidth]{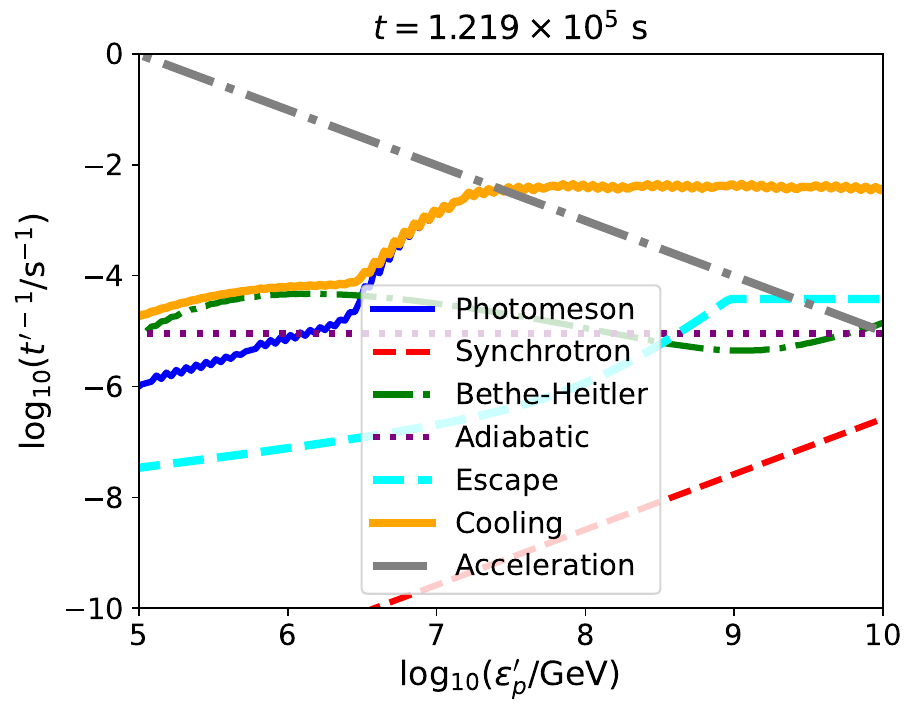}
\includegraphics[width=0.32\linewidth]{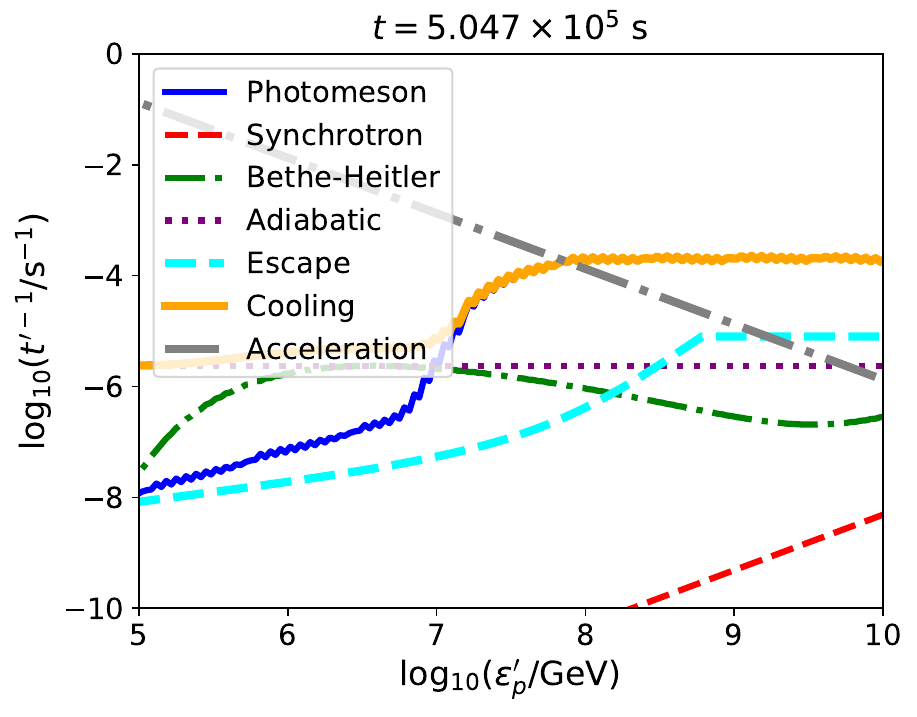}
\includegraphics[width=0.32\linewidth]{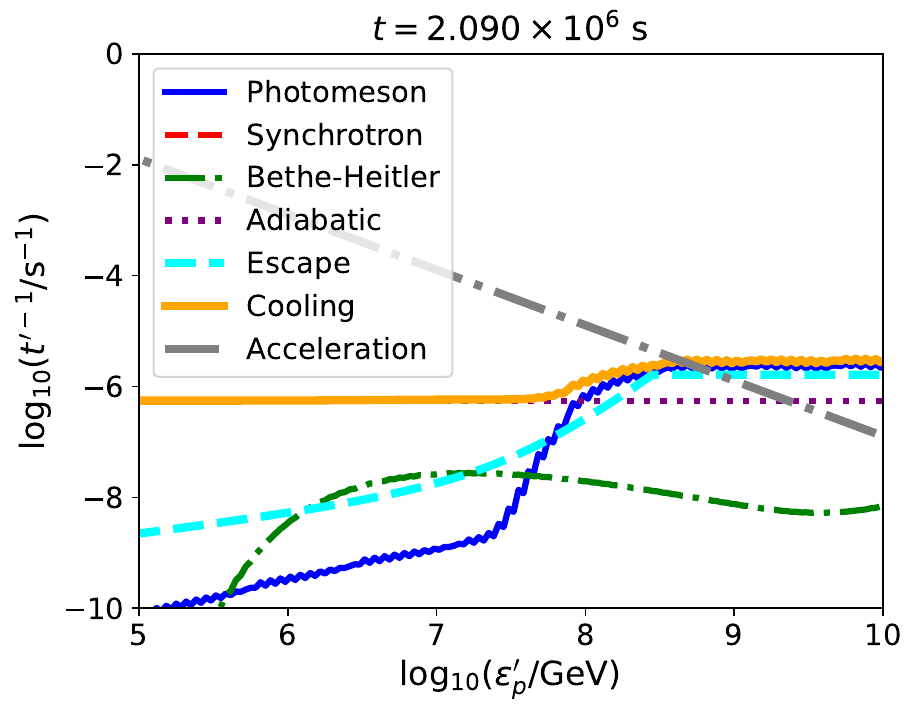}\\
\includegraphics[width=0.32\linewidth]{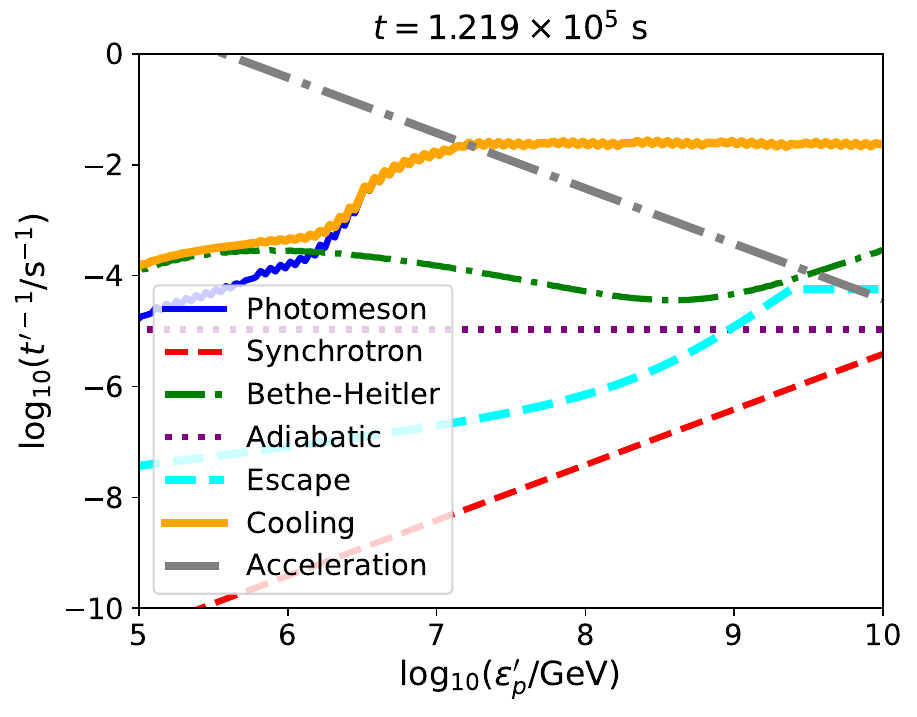}
\includegraphics[width=0.32\linewidth]{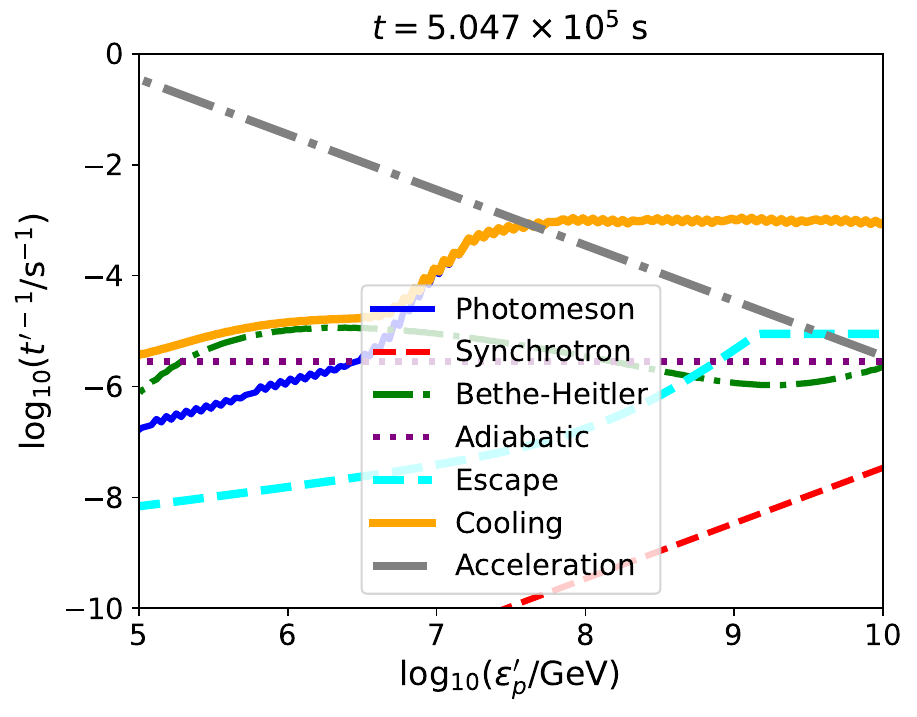}
\includegraphics[width=0.32\linewidth]{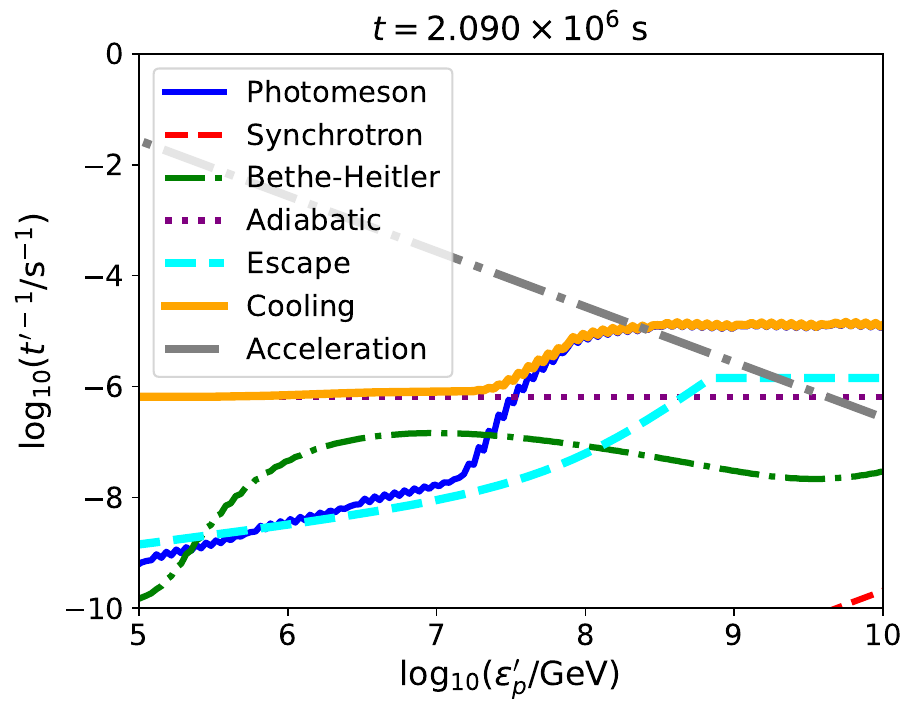}
\caption{\label{fig:timescales} Plots showing the rates of various acceleration and cooling processes with comoving proton energy $\varepsilon^\prime_p$ at various timesnaps. The \emph{upper} panel shows the \emph{fiducial} case, while the \emph{lower} panel shows the \emph{optimistic} scenario. Refer to Table~\ref{tab:params} for details on the choice of parameters for both scenarios.
}
\end{center}
\end{figure*}
In this subsection, we focus on the various relevant timescales in the context of neutrino production. We begin with the the energy loss timescale ($t_{\rm loss}^\prime $). The loss rate is given by $t_{\rm loss}^{\prime -1}=t_{\rm esc}^{\prime  -1}+t_{\rm cool}^{\prime -1}$, where $t_{\rm esc}^\prime $ and $t_{\rm cool}^\prime $ are the escape and the cooling timescales for the protons, respectively. The escape timescale for the protons is given by $t_{\rm esc}^\prime  = {\rm max}\left[ R(t)^2/D_c(\varepsilon_p^\prime), R(t)/c \right]$, where the second term is just the light crossing time. The diffusion coefficient $D_c$ is given by~\citep{Harari:2013pea},
\begin{equation}
\label{eq:crdiffusion}
D_c(\varepsilon_p^\prime ) = \frac{c}{3} l_c \left[ 4 \left( \frac{\varepsilon_p^\prime }{\varepsilon_c^\prime } \right)^2 + a_I \left( \frac{\varepsilon_p^\prime }{\varepsilon_c^\prime } \right)+ a_L \left( \frac{\varepsilon_p^\prime }{\varepsilon_c^\prime } \right)^{(2-m) } \right]\,.
\end{equation}
We assume a Kolmogorov cascade for the CR diffusion, for which $m = 5/3$. For this choice of $m$, the coherence length of the turbulent magnetic field is given by, $l_c^\prime  =  R(t)/\big( \Gamma_{\rm ej} 5\big)$ and numerical fits fix the coefficients $a_I \approx 0.9$ and $a_L \approx 0.23$. The CR diffusion timescale is characterized by a critical energy ($\varepsilon_c^\prime $) such that the effective Larmor radius $r_L^\prime  = \varepsilon_p^\prime / \big( Ze B_d \big)$ satisfies, $r_L^\prime (\varepsilon_c^\prime ) = l_c^\prime $. This defines $\varepsilon_c^\prime  = l_c^\prime  Z e B_d $.

The cooling rate for the CR protons is given by\footnote{Note that since proton densities in the nebula is sufficiently low, we can ignore $pp$ interactions for this system. Besides the kilonova ejecta also has low densities at the relevant timescales for proton escape. Hence neutrino productions from $pp-$interactions are not efficient so we do not take those into account.} $t_{\rm cool}^{\prime -1} =  t_{\rm p\gamma}^{\prime -1} + t_{\rm sync}^{\prime -1} + t_{\rm BH}^{\prime -1} + t_{\rm dyn}^{\prime -1}$, where $t_{pp}^\prime , t_{\rm p\gamma}^\prime , t_{\rm sync}^\prime , t_{\rm BH}^\prime , t_{\rm dyn}^\prime $ denote $pp$, $p\gamma$, proton synchrotron (sync), Bethe-Heitler (BH), and dynamical (dyn) cooling timescales respectively. The proton synchrotron timescale $t_{\rm sync}^\prime  = 6 \pi m_p^4 c^3/\big( m_e^2 \sigma_T \varepsilon_p^\prime B_{\rm neb}^{\prime 2}\big)$, where $\sigma_T$ is the Thomson cross-section. The dynamical timescale is defined as $t_{\rm dyn}^\prime  = R/\big( \Gamma_{\rm ej} v \big)$. The $p\gamma$ and BH cooling rates can be computed using the following equation:
\begin{equation}
t_{p\gamma}^{\prime -1} (\varepsilon_p^\prime) = \frac{c}{2 \gamma_p^2} \int_{\bar{\varepsilon}_{\rm th}}^{\infty} d \bar{\varepsilon}_\gamma\ \kappa (\bar{\varepsilon}_\gamma ) \sigma (\bar{\varepsilon}_\gamma )\bar{\varepsilon}_\gamma \int_{\bar{\varepsilon}_\gamma /2\gamma_p}^{\infty} d \varepsilon_\gamma^\prime \varepsilon_\gamma^{\prime -2} \frac{d N_\gamma^\prime}{d \varepsilon_\gamma^\prime}\,,
\end{equation}
where the Lorentz factor of the CR protons is given by $\gamma_p = \varepsilon_p^\prime/\big( m_p c^2 \big)$, the threshold energy, inelasticity, and cross-sections in the rest frame of the CRs is given by $\bar{\varepsilon}_{\rm th}$, $\sigma (\bar{\varepsilon}_\gamma )$, and $\kappa (\bar{\varepsilon}_\gamma )$ respectively, for the $p\gamma$ and BH processes. The energy of the photon in the comoving frame is given by $\varepsilon_\gamma^{\prime}$ and the energy of the photon in the comoving frame of the proton is given by $\bar{\varepsilon}_\gamma$. The cross-section and the inelasticity for the $p\gamma$ process is used from~\cite{Murase:2006dr} and for the BH process the same is done from~\cite{1983MNRAS.204.1269S,1992ApJ...400..181C}. The photon spectra $d N_\gamma^\prime/d \varepsilon_\gamma^\prime$ is briefly discussed in Section~\ref{subsec:elec_phot_spectra}.

The various acceleration and cooling rates of the protons for the fiducial (optimistic) case, as a function of their energy, at different timesnaps are shown in Figure~\ref{fig:timescales} in the upper (lower) panel. We note that the rates at various timesnaps for both the fiducial and the optimistic cases are approximately the same. At initial times ($t \sim 10^5$ s), the rate of acceleration of protons at the TS dominates over the cooling rate for $\varepsilon_p^\prime \leq 10^7$ GeV. Furthermore, for protons of energy $\varepsilon_p^\prime \gtrsim 10^7\ {\rm GeV}$ the photomeson cooling dominates the cooling rate. During this time, the target photons for the protons have energies\footnote{
This can be estimated assuming the $\Delta$-resonance process, using $\varepsilon_\gamma^\prime \approx (\bar{\varepsilon}_{\rm pk}^\prime/\varepsilon_p^\prime) (m_p c^2)$, where the photon energy at the $\Delta$-resonance peak is $0.3$ GeV.}
$\lesssim 30$ eV. The temperature of the thermal photons can be estimated using $T_{\rm th} \sim \left( E_{\rm th}/( a\ 4 \pi R^2 \Gamma_{\rm ej}^2 \beta_{\rm ej} c t ) \right)^{1/4} \sim 6.4 \times 10^4$ K, where $a$ is the radiation constant ($E_{\rm th}$ can be estimated from Figure~\ref{fig:energies}). The resultant thermal photon energy $\varepsilon_\gamma^{\prime \rm th} \sim 2.8\ k_B T_{\rm th} \sim 10$ eV, where $k_B$ is the Stefan-Boltzmann constant ($k_B = ac/4$). Since $\varepsilon_\gamma^{^\prime \rm th} < 30$ eV, the thermal photons cannot serve as target for the protons for meson production. This implies that the target photons for the protons during this time have a non-thermal spectrum. For protons with energies $\varepsilon_p^\prime < 10^7\ {\rm GeV}$ that interact with thermal photons, the Bethe-Heitler cooling is efficient. 

For an intermediate timesnap associated with significant neutrino production ($t \sim 5 \times 10^5$ s), photomeson cooling is efficient for $\varepsilon_p^\prime \gtrsim 10^7\ {\rm GeV}$. For late times in the context of neutrino production ($t \sim 10^6$ s), the rate of proton acceleration is efficient across all proton energies $\lesssim 10^{8.5}$ GeV, but less than the intermediate timesnap $t \sim 5\times 10^5$ s. The cooling rate is dominated by adiabatic cooling and photomeson cooling only dominates for proton energies $> 10^8$ GeV. The target photon spectra during the intermediate and late times are also the non-thermal photons in the nebula, since $T_{\rm th}$ drops rapidly for $t>t_{\rm sd}$. Furthermore, it is also interesting to note that for $\varepsilon_p^\prime \gtrsim 10^8$ GeV the proton escape rate is comparable to the cooling rate for the fiducial case.

The TS region can also accelerate ions due to shocks or magnetic reconnection~\citep{Sironi:2011zf,Sironi:2014jfa}. Thus following the polar cap acceleration a fraction of the protons can be accelerated at the TS region to higher energies up to $\varepsilon_p^{\prime \rm cutoff,TS}$ with a spectral index close to $-2$, where $\varepsilon_p^{\prime \rm cutoff,TS}$ is the cut-off energy in the TS region. This can be calculated by balancing the energy loss timescale ($t_{\rm loss}^\prime $) and the acceleration ($t_{\rm acc}^\prime $) timescale for the CR protons. The acceleration timescale is given by $t_{\rm acc}^{\prime} = \eta_{\rm acc}\varepsilon_p^\prime/\big(Z e c B_{\rm neb}^\prime \big)$, where $B_{\rm neb}^\prime$ is the co-moving magnetic field from Equation~\eqref{eq:bneb}, $\eta_{\rm acc}$ gives the efficiency of the acceleration. In general, $\eta_{\rm acc} \sim 1 - 10$, so we choose $\eta_{\rm acc} = 1$ for this work.

However, for our choice of parameters, the TS acceleration plays no role at all for the relevant timescales, since, at all times $\varepsilon_p^{\prime \rm cutoff, pc} > \varepsilon_p^{\prime \rm cutoff,TS}$. This is a combination of factors including the choice of a comparatively low value for $\epsilon_B = 10^{-4}$ and a higher photon density as a result of the electron injection spectra (see Section~\ref{subsec:elec_phot_spectra}). This is also evident from Figure~\ref{fig:picooling} (left panel)  where $\varepsilon_p^{\prime \rm cutoff, TS}$ (solid) is shown for the fiducial (thick red-orange) and the optimistic (thin purple) scenarios. The cut-off energy increases with time, peaks for $t> t_{\rm sd}$, and decreases following the peak.
\subsection{Neutrino spectra}
In this sub-section, we discuss the processes contributing to the neutrino spectra. In our model, protons interact with photons in the nebula, which mainly produces pions. These pions decay to neutrinos and muons, and the muons decay to neutrinos and electrons/positrons. To calculate the neutrino spectra, we use the fitting formula given by \cite{Kelner:2008ke}:
\begin{equation}
N_{\varepsilon_l}^{\rm KA}\approx \int d \varepsilon_\gamma^\prime \frac{d\varepsilon_p^\prime}{\varepsilon_p^\prime} \frac{d N_p}{d\varepsilon_p^\prime} \frac{d N_\gamma^\prime}{d\varepsilon_\gamma^\prime}
\Phi_l \left( \eta, \frac{\varepsilon_\gamma^\prime}{\varepsilon_p^\prime} \right) \,,
\end{equation}
where $\Phi_l(\eta,\varepsilon_\gamma^\prime/\varepsilon_p^\prime)$ is a function that includes the differential cross-section based on simulations using the code SOPHIA \citep{2000CoPhC.124..290M}, $l$ is a lepton species, and $\eta=4\varepsilon_\gamma^\prime\varepsilon_p^\prime/(m_p^2c^4)$. We also take into account the suppression factor associated with cooling of pions and muons. These are given by $f_{i,{\rm sup}} = 1 - {\rm exp} \big( -t^\prime_{i,\rm cool}/t^\prime_{i, \rm dec} \big)$, where the comoving cooling rates and the decay timescales of particle species $i$ are given by $t^{\prime -1}_{i,\rm cool} \approx t^{\prime -1}_{i,\rm sync} + t^{\prime -1}_{\rm dyn}$ and $t^\prime_{i,\rm dec} = t_i \varepsilon_i^\prime/\big( m_i c^2 \big)$, respectively. 
The comoving synchrotron timescale $t^\prime_{i,\rm sync} = 6 \pi m_i^4 c^3/\big( m_e^2 \sigma_T \varepsilon_i^\prime B_{\rm neb}^{\prime 2}\big)$, $t_i$ is the lifetime of $i$ in the rest frame. We approximately estimate the neutrino spectra for muon-neutrino and electron-neutrino to be
\begin{align}
\label{eq:nuspec}
N_{\varepsilon_{\nu_e+\bar{\nu_e}}} &\approx\left(N_{\varepsilon_{\nu_e}}^{\rm KA}+N_{\varepsilon_{\bar{\nu_e}}}^{\rm KA} \right)f_{\mu,\rm sup}f_{\pi,\rm sup}, \\
N_{\varepsilon_{\nu_\mu+\bar{\nu_\mu}}} &\approx \left(N_{\varepsilon_{\nu_\mu}}^{\rm KA}+N_{\varepsilon_{\bar{\nu_\mu}}}^{\rm KA}- N_{\varepsilon_{e^+}}^{\rm KA} - N_{\varepsilon_{e^-}}^{\rm KA}\right)f_{\pi,\rm sup} + N_{\varepsilon_{\nu_e+\bar{\nu_e}}}\,,
\end{align}
where the first term in the parentheses accounts for pion decay ($\nu_\mu,\bar{\nu}_\mu,\mu^+,\mu^-$) and the second term accounts for the products from muon decay ($\nu_e,\bar{\nu}_e,\nu_\mu,\bar{\nu}_\mu,e^+,e^-$). Neutrinos are produced by the decay of muons and pions. Electron neutrinos are produced only by muon decay, and muon neutrinos are produced by both pion and muon decays. The formula for muon neutrinos given by~\cite{Kelner:2008ke} contains both the pion- and muon-decay components. Thus, we need to subtract the muon-decay component from the muon-neutrinos ($N_{\epsilon_\mu}^{\rm KA}$) in order to estimate the pion-decay component. Since electron neutrinos are produced only by muon decay, we approximately treat $N_{\epsilon_e}^{\rm KA}$ as the muon-decay component, that is, $N_{\epsilon_\mu}^{\rm KA} \approx N_{\epsilon_e}^{\rm KA} \approx N_{\epsilon_{e^+}}^{\rm KA} + N_{\epsilon_{e^-}}^{\rm KA}$. The neutrino flux without taking into account the neutrino oscillation is given by $\phi_i^0 = N_{\varepsilon_i} /\big( 4 \pi d_L^2 \big)$, where $i$ denotes the neutrino species and $d_L$ is the luminosity distance of the source. The neutrino oscillations change the flavor ratio during the neutrinos propagation from the source to the Earth. The oscillated flux on Earth is given by~\citep{BECKER2008173}
\begin{align}
\label{eq:flux}
\phi_{\nu_e + \bar{\nu}_e} &= \frac{10}{18} \phi_{\nu_e + \bar{\nu}_e}^0 + \frac{4}{18} \left( \phi_{\nu_\mu + \bar{\nu}_\mu}^0 + \phi_{\nu_\tau + \bar{\nu}_\tau}^0 \right)\,,\nonumber\\
\phi_{\nu_\mu + \bar{\nu}_\mu} &= \frac{4}{18} \phi_{\nu_e + \bar{\nu}_e}^0 + \frac{7}{18} \left( \phi_{\nu_\mu + \bar{\nu}_\mu}^0 + \phi_{\nu_\tau + \bar{\nu}_\tau}^0 \right)\,.
\end{align}
The pion-suppression factor ($f_{\pi,\rm sup}$) appearing in Equation~\eqref{eq:nuspec} plays an important role in deciding the neutrino fluences and hence it is important to understand the regimes where pion suppression becomes important. In photomeson cooling, the parent proton with energy $\varepsilon_p^\prime$ produces pions with mass $m_\pi$ and energy $\varepsilon_\pi^\prime = \varepsilon_p^\prime/5$ that then decay to produce neutrinos. However, the neutrino production can be suppressed if instead pions cool through synchrotron cooling before they can decay to neutrinos. Thus pion suppression becomes important in these scenarios. In Figure~\ref{fig:picooling} we show the time evolution of the cut-off energy for the protons in the polar cap region $\varepsilon_{\rm max}^{\prime \rm pc}$ (solid) (see Equation~\ref{eq:epcutoffpc}) and the threshold pion energy $\varepsilon_{\pi, \rm th}^\prime$ (dashed) in terms of the proton energy $\varepsilon_{p, \rm th}^\prime$ (that is scaled by a factor $5$) for the fiducial (thick red-orange) and optimistic (thin purple) cases. The threshold energy for pion cooling is computed by equating the synchrotron cooling timescale ($t_{\pi, \rm sync}^\prime = 6 \pi m_{\pi}^4 c^3/\big( m_e^2 \sigma_T \varepsilon_\pi^\prime B_{\rm neb}^{\prime 2}\big)$) to the pion decay timescale ($t_{\pi,\rm dec}^\prime = \varepsilon_{\pi}^\prime \tau_{\pi}/\big(m_{\pi} c^2\big)$). This allows us to compute the threshold in terms of the proton energy $\varepsilon_{p,\rm th}^\prime$ as,
\begin{equation}
\label{eq:epithreshold}
\varepsilon_{p,\rm th}^\prime = 5\varepsilon_{\pi, \rm th}^\prime = 5 \sqrt{\frac{6 \pi m_\pi^5 c^5}{m_e^2 \sigma_T \tau_\pi}} \frac{1}{B^\prime_{\rm neb}}\,,
\end{equation}
where $B_{\rm neb}^\prime$ is given by Equation~\eqref{eq:bneb} and the decay time of charged pions at rest is given by $\tau_\pi \approx 2.6 \times 10^{-8}$ s. For protons with energy $\varepsilon_p^\prime > \varepsilon_{p,\rm th}^\prime$, pion cooling is important and suppresses the neutrino production. This is because in this case the pion decay time is longer and synchrotron cooling is faster for higher energy pions.

From Figure~\ref{fig:picooling}, we see that for $t \lesssim 10^4$ s, $\varepsilon_{p,\rm th}^\prime < \varepsilon_{p}^{\prime \rm cutoff,pc}$, which marks the regime when pion suppression is indeed important. Defining $f_{\pi, \rm sup} \approx 1 - {\rm exp} \big( -t_{\pi, \rm sync}^\prime/t_{\pi, \rm dec}^\prime \big)$, in this regime we have $f_{\pi, \rm sup} < 1$ leading to a suppression of neutrino production. For timescales $t>10^4$ s, $\varepsilon_{p,\rm th}^\prime > \varepsilon_{p}^{\prime \rm cutoff,pc}$, implying $f_{\pi, \rm sup} \sim 1$ and hence pion suppression has no effect on the neutrino production.
\section{Neutrino fluences and detectability}
\label{sec:res}
In this section we focus on the neutrino fluences expected from our model, the contribution of neutrinos from magnetar remnants from BNS mergers to the diffuse high-energy neutrino flux, and the prospects of GW-triggered neutrino searches to detect the predicted high-energy neutrino emission from our magnetar model.
\subsection{Neutrinos from individual sources}
\label{sec:nu_individual}
\begin{figure*}
\begin{center}
\includegraphics[width=0.48\textwidth]{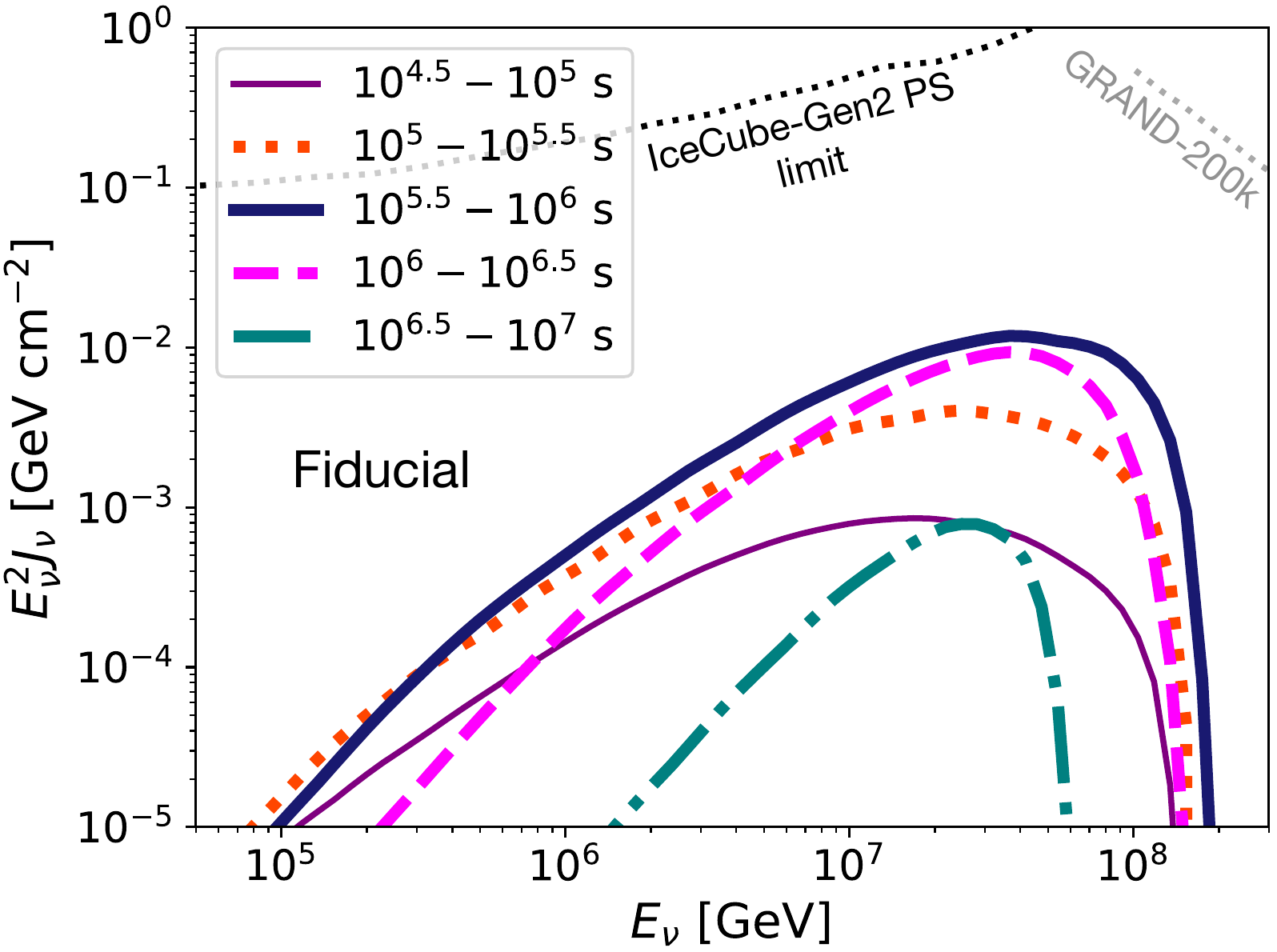}
\includegraphics[width=0.48\textwidth]{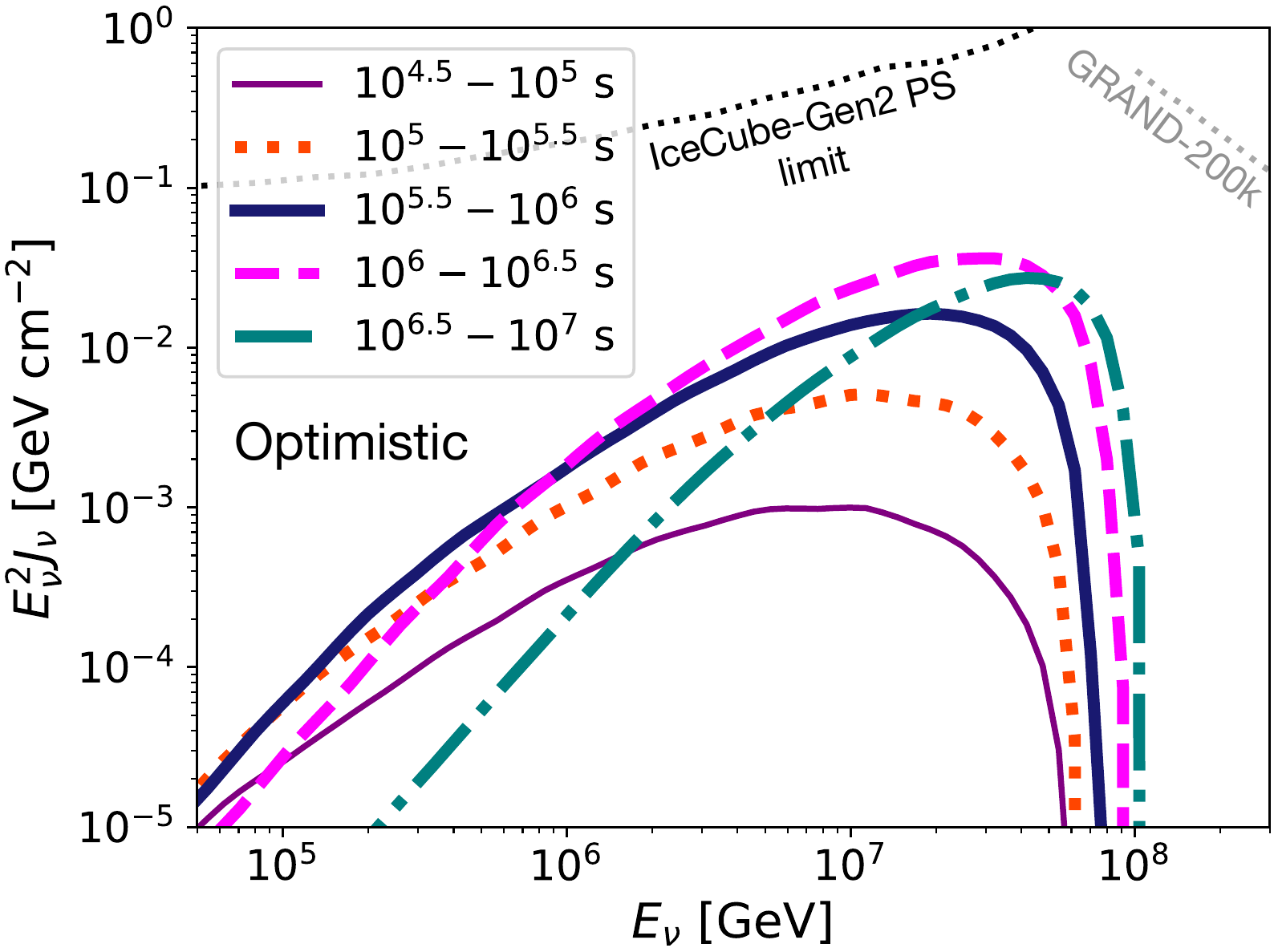}
\caption{\label{fig:nu}High-energy neutrino fluence (all-flavors) from a magnetar at $40$ Mpc for various timescales ranging from a few hours to a few days post the merger. The \emph{fiducial} and the \emph{optimistic} scenarios are shown in the \emph{left} and \emph{right} panels respectively. Refer to Table~\ref{tab:params} for details on the choice of parameters for both scenarios.
}
\end{center}
\end{figure*}
Recall that we discussed in Section~\ref{subsec:cr_prot} that the CR protons can be accelerated at the polar cap region, followed by a fraction of them being reaccelerated at the region of termination shock (TS). Let us first try to qualitatively understand the characteristics of the neutrino fluences in these two regions. Based on our choice of parameters both for the fiducial and optimistic cases, the acceleration in the TS region is not effective since $\varepsilon_p^{\prime \rm cutoff,pc} > \varepsilon_p^{\prime \rm cutoff,TS}$ at all times. This can also be noticed from the \emph{left panel} of Figure~\ref{fig:picooling}.

The neutrino fluences resulting from the acceleration across the polar cap region are shown in Figure~\ref{fig:nu} for various epochs of evolution of the magnetar system, for a source located at luminosity distance $d_L = 40$ Mpc (similar to GW170817). The parameters for the fiducial and optimistic cases are shown in Table~\ref{tab:params}. The neutrino fluence for the fiducial case peaks between $10^7 {\rm GeV} - 10^8 {\rm GeV}$ and is $\sim 1 \times 10^{-2} {\rm GeV}\ {\rm cm}^{-2}$. This is lower than that of the optimistic case where the neutrino fluence peaks around similar energies but is $\sim 4 \times 10^{-2} {\rm GeV}\ {\rm cm}^{-2}$. The difference mainly stems from the larger energy in protons ($E_p$) as a result of the higher spindown energy ($E_{\rm sd}$) owing to the smaller initial spin period ($P_i = 1$ ms) in the optimistic case (see Figure~\ref{fig:energies}). The neutrino fluences can be estimated from the parent proton energies by~\citep{Kimura:2022zyg}, 
\begin{equation}
E_\nu^2 J_\nu \sim \frac{1}{4 \pi d_L^2} \left(\frac{3}{8}E_p  f_{p\gamma} f_{\rm bol} f_{\pi, {\rm sup}}\right)\,,
\end{equation}
where $f_{p\gamma}$ is defined as the efficiency of pion production, that is, the fraction of CR protons producing pions through the photomeson process, $f_{\rm bol}$ is defined as the bolometric correction factor which gives the fraction of the total energy of CR protons that goes into neutrino production.
From Figure~\ref{fig:energies}, we see that the total proton energy around the time of the peak neutrino fluence, $10^5$ s ($10^{5.5}$ s), is roughly $E_p \sim 2 \times 10^{49}$ erg ($1.8 \times 10^{50}$ erg) for the fiducial (optimistic) case. From Figures~\ref{fig:timescales} and~\ref{fig:picooling}, we can deduce that at the time of the neutrino fluence peaks $f_{p\gamma}$ and $f_{\pi, {\rm sup}}$ are roughly $0.8 - 1$ and unity respectively. 
The bolometric factor 
$f_{\rm bol} \sim \varepsilon_p^{\prime 2} d\dot{N}_p/d\varepsilon_p^\prime|_{\varepsilon_p^\prime = \varepsilon_p^{\prime,\rm cutoff, pc}}/\big( \int d\varepsilon_p^\prime \varepsilon_p^\prime d\dot{N}_p/d\varepsilon_p^\prime \big)$ can be estimated to be $\sim 1.0$ around the time of the peak neutrino fluences. 
Putting it altogether we have, $E_\nu^2 J_\nu \sim 2 \times 10^{-2}{\rm GeV}\ {\rm cm}^{-2}\ (1 \times 10^{-1}{\rm GeV}\ {\rm cm}^{-2})$ for the fiducial (optimistic) case. This is within a factor of a few ($2-4$) as compared to what we obtained from the numerical results. This discrepancy mainly stems from the energy spectrum of the secondary particles that is taken into account only in the numerical results. Besides, for the analytical estimates instead of integrating over time, we multiply with time which introduces some discrepancies. 

One of the most important takeaways from Figure~\ref{fig:nu} is the optimal time window to search for high-energy neutrinos from these magnetar remnants of BNS mergers is $\sim 10^{5.5} - 10^{6.5}$ s, that is, $\mathcal{O}(1 - 10\ \rm days)$ post-merger depending on the fiducial or optimistic case. The peak energy of the neutrino spectra and the cutoff energy is decided by the cooling and acceleration rates of the protons along with $f_{p\gamma}, f_{\pi,\rm sup},$ and $f_{\rm bol}$. At the initial times, the neutrino fluence is suppressed due to pion cooling. This can be seen from Figure~\ref{fig:picooling} (\emph{right panel}), where we note that pion cooling dominates for timescales $\lesssim 10^4$ s and thus $f_{\pi, \rm sup} <1$. For the initial times we see that the neutrino fluence peaks around $\sim {\rm a\ few}\ 10^7$ GeV. This is easy to see from Figure~\ref{fig:timescales} (panels corresponding to $t\sim 10^5$ s), where the photomeson cooling efficiency dominates the cooling channel around that energy scale and higher in proton energy. The peak energy of the neutrino fluence increases as the system evolves in time. At very late times the peak energy saturates at values $\lesssim 10^8$ GeV.

There are two very crucial physical parameters for neutrino production: the initial spindown time ($P_i$) and the dipolar magnetic field strength of the magnetar ($B_d$). A lower value of $P_i$ gives a higher neutrino fluence. This is reasonable since a shorter initial spin down time leads to a higher spindown energy, thus providing a larger energy source for neutrino production as compared to the fiducial case (see Figure~\ref{fig:energies}). As the $P_i$ increases the neutrino fluxes go down leading to less optimistic scenarios. The neutrino fluences have a non-monotonic behaviour with respect to the strength of the dipolar magnetic field ($B_d$). We scanned the parameter space between $B_d = 10^{13} - 10^{15}$ G and found that the neutrino production peaks around $B_d \sim 2.5 \times 10^{13}$ G. This is due to a combination of several factors like the number of accelerated protons and the efficiency of photomeson cooling for example. Decreasing $B_d$ to $10^{13}$ G (compared to $10^{14}$ G in the fiducial case) leads to a drop in the neutrino flux. The parameters $M_{\rm ej}$ and $v_0$ does not play a significant role in changing the neutrino fluxes. This can be understood from the fact that the mass of the ejecta and how fast it is moving has no direct effects on the neutrino production processes for our choice of parameters. In general, the mass and velocity of the expanding ejecta can affect the thermal radiation energy density which in turn would have an effect on the neutrino fluence. However, as discussed in Section~\ref{subsec:timescales}, for our choice of parameters, the neutrino production occurs due to photomeson cooling where the targets are the non-thermal photons and hence our results are mostly unaffected by the choice of $v$ and $M_{\rm ej}$.

We also show the point source sensitivities for IceCube-Gen2 (dotted black curve) and GRAND-200k (dotted gray curve) in Figure~\ref{fig:nu}. For the former we take the current point source limit for IC-86 (IV) between declination bin $-2.29^\circ - 0^\circ$ and multiply it by $10^{2/3}$ to scale for Gen2. For the latter, we use the effective area corresponding to sensitivity at zenith angle $90^\circ$. Given the declination dependent effective area ($A_{\rm eff} (\delta, \varepsilon_{\nu_\mu})$), the number of muon neutrinos (and antineutrinos) in a detector is given by 
\begin{equation}
\label{eq:nmunu}
N_{\nu_\mu+\bar{\nu}_\mu} = \int d\varepsilon_{\nu_\mu} \phi_{\nu_\mu + \bar{\nu}_\mu} \big( \varepsilon_{\nu_\mu} \big) A_{\rm eff}\big( \delta, \varepsilon_{\nu_\mu} \big)\,.
\end{equation}
The number of $\nu_\mu + \bar{\nu}_\mu$ neutrinos for a typical source at $40$ Mpc can then be  given by $N_{\nu_\mu+\bar{\nu}_\mu} \sim 0.1\ (0.4)$ for the fiducial (optimistic cases), assuming the effective area corresponding to the projected IceCube-Gen2 limit shown in Figure~\ref{fig:nu}.
Thus it is evident that for a typical source at $40$ Mpc, the prospects for individual source detections are low. This paves the way for looking at high-energy neutrino detections from magnetar sources in the diffuse neutrino channels and the prospects of performing GW-triggered stacking searches.
\subsection{Diffuse Neutrinos}
\begin{figure}
\includegraphics[width=0.48\textwidth]{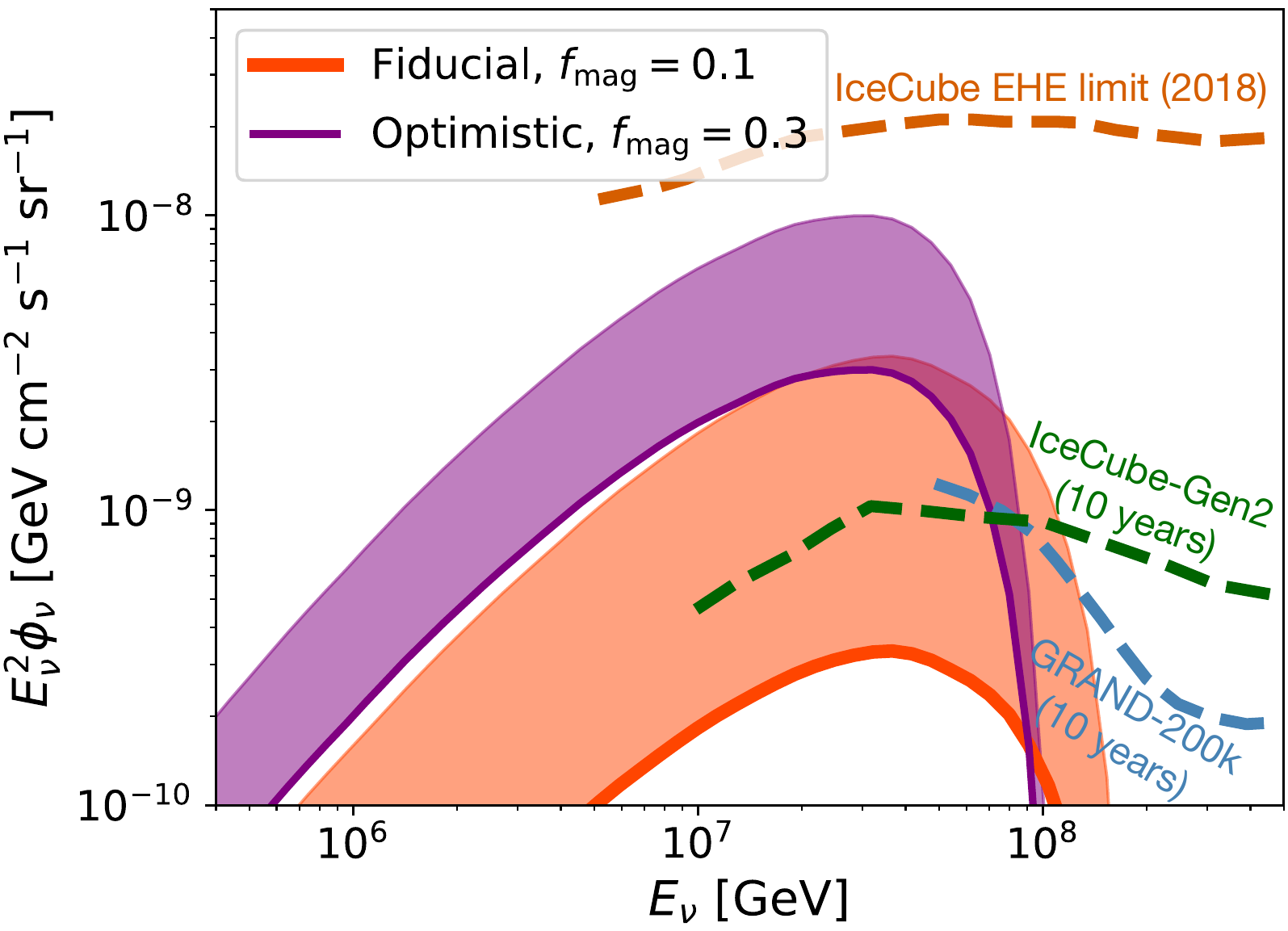}
\caption{\label{fig:diffuse_nu} Diffuse high-energy neutrino flux (all-flavors) from magnetar sources for the fiducial (thick red orange line) and optimistic (thin purple line) scenarios assuming the fraction of BNS mergers forming stable magnetars $f_{\rm mag}= 0.1$ and 0.3 respectively. The region between $f_{\rm mag}^{\rm fid}$ or $f_{\rm mag}^{\rm opt}$ and the most-optimistic value $f_{\rm mag} = 1.0$ is shaded. The $90$\% C.L. for the all-flavor differential flux from the 9 year IceCube data is shown as a dotted dark orange line~\citep{IceCube:2018fhm}. The 10-year projected sensitivities for all-flavor diffuse flux for IceCube-Gen2 (optical+radio) and GRAND-200k is shown in dashed dark green and steel blue lines respectively~\citep{Gen2_TDR}. Refer to Table~\ref{tab:params} for details on the choice of parameters for both scenarios.
}
\end{figure}
In Section~\ref{sec:nu_individual} we found that the neutrino fluences from individual sources at typical distances of $\mathcal{O}(10)$ Mpc may not be detectable in the upcoming neutrino detectors. We estimate the diffuse high-energy neutrino fluxes from magnetar remnants from BNS mergers and find that the magnetar model cannot contribute significantly to the diffuse neutrino flux. The diffuse neutrino flux can be roughly estimated using
\begin{equation}
\label{eq:diffusenu}
E_\nu^2 \phi_\nu = \frac{c}{4 \pi H_0} \bigg( E_\nu^2 \frac{dN_\nu}{dE_\nu} \bigg) \big( f_{\rm mag} \dot{\rho}(z=0)\big) f_z\,,
\end{equation}
where $H_0$ is the Hubble constant $H_0 = 72\ {\rm km\ s}^{-1}{\rm Mpc}^{-1}$, $\dot{\rho} (z=0)$ gives the local rate of BNS mergers at redshift $z = 0$, and $f_z$ is the redshift factor. We choose, $\dot{\rho} (z=0) = 300\ {\rm Gpc}^{-3} {\rm yr}^{-1}$~\citep{KAGRA:2021duu} and $f_z = 2$. Finally, $f_{\rm mag}$ gives the fraction of BNS mergers that form long-lived magnetar remnants. It is unclear what fraction of BNS mergers result in stable magnetar like remnants, and hence $f_{\rm mag}$ has a large uncertainty~\citep{Wang:2023qww}. We choose a conservative value for $f_{\rm mag}$ and set it to $0.1$ and $0.3$ for the fiducial and optimistic scenarios respectively (note that this choice is ad-hoc and hgihlights two representative cases). Evidently, choosing the best possible value of $f_{\rm mag} = 1$ will yield very optimistic results for our model. Also note that Equation~\ref{eq:diffusenu} is a simple expression to estimate the diffuse flux, which is sufficient for our current work. This gives a peak diffuse neutrino flux of $\sim 2 \times 10^{-10} {\rm GeV}{\rm cm}^{-2} {\rm s}^{-1} {\rm sr}^{-1}$ and $\sim 2 \times 10^{-9} {\rm GeV}{\rm cm}^{-2} {\rm s}^{-1} {\rm sr}^{-1}$ between $10^7 - 10^8$ GeV for the fiducial and the optimistic cases respectively. 

We show the diffuse high-energy neutrino flux from magnetar sources in Figure~\ref{fig:diffuse_nu}. The thick orange-red solid and thin purple lines show the results from the fiducial and optimistic cases with $f_{\rm mag}^{\rm fid} = 0.1$ and $f_{\rm mag}^{\rm opt} = 0.3$ respectively. We also shade the region extending from $f_{\rm mag}^{\rm fid}$ or $f_{\rm mag}^{\rm opt}$ to $f_{\rm mag} = 1$ to highlight the most optimistic result possible. For the realistic choices of $f_{\rm mag}$ both the fiducial and optimistic scenarios are below the current IceCube EHE limit~\citep{IceCube:2018fhm}. However the optimistic scenario even with a conservative choice of $f_{\rm mag}^{\rm opt}$, lies within the projected 10-year sensitivity of IceCube-Gen2 (optical and radio)~\citep{Gen2_TDR} and GRAND-200k~\citep{GRAND:2018iaj}, but the detection potential of the latter is less optimistic due to the relevant energy range being lower than the peak sensitivity of GRAND-200k. For the fiducial case, one can expect a detection if $f_{\rm mag}^{\rm opt} \gtrsim 0.4$.
\subsection{Neutrino-GW associations}
\begin{figure*}
\centering
\includegraphics[width=\textwidth]{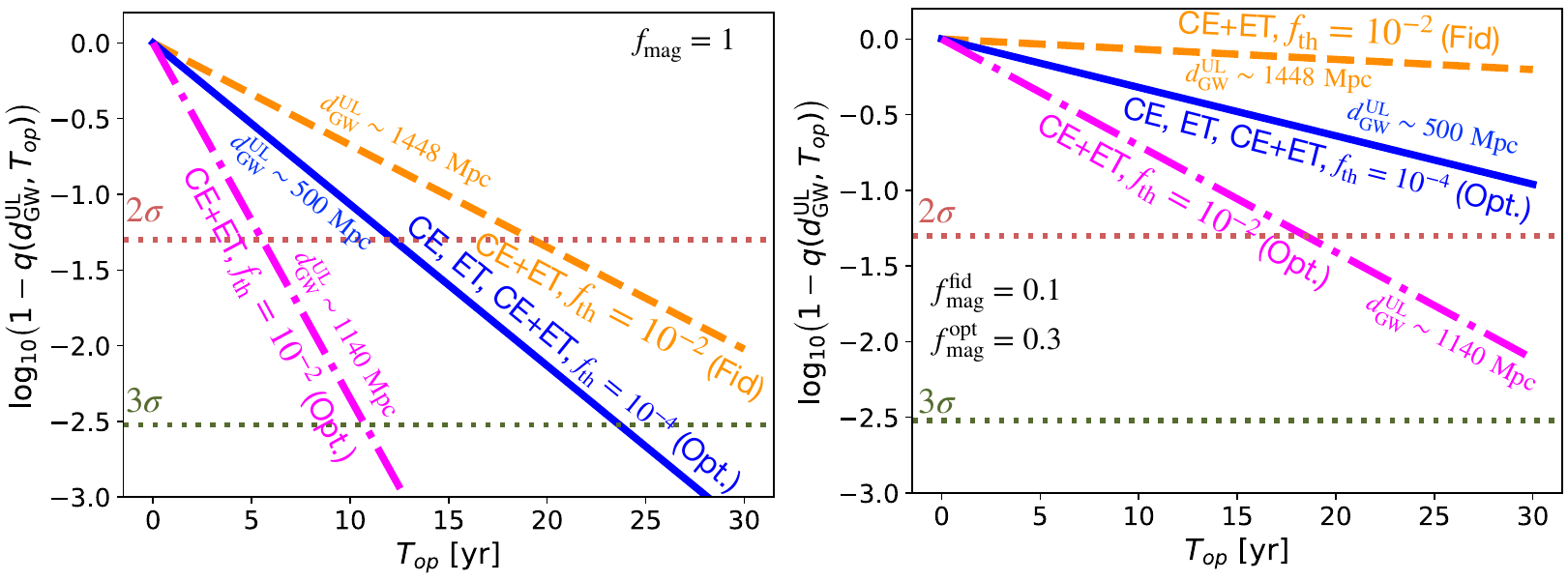}
\caption{\label{fig:nu_gw} Probability of neutrino detection ($q$) with operation time $T_{\rm op}$ for gravitational-wave triggered searches at IceCube-Gen2, for high-energy neutrinos from magnetar remnants from BNS mergers, using the next-generation GW detectors - the Cosmic Explorer (CE), the Einstein Telescope (ET), and the combination of the two: CE+ET - for the most-optimistic value of $f_{\rm mag} = 1$ (\emph{left panel}) and for realistic values of $f_{\rm mag}^{\rm fid} = 0.1$ and $f_{\rm mag}^{\rm opt} = 0.3$ (\emph{right panel}). The analysis takes into account the use of localization capabilities of the next-generation detectors to obtain a limiting distance for the GW detectors ($d_{\rm GW}^{\rm UL}$)~\citep{Mukhopadhyay:2023niv,Mukhopadhyay:2024lwq}. The $2\sigma$ and $3\sigma$ C.L.s are also shown with horizontal dashed lines. The parameters used for the fiducial (shown as the dashed orange line) and the optimistic (shown as the solid blue and dot-dashed pink lines) scenarios are given in Table~\ref{tab:params}.
}
\end{figure*}
In the previous sections we discussed the prospects of detecting high-energy neutrinos are less optimistic. To alleviate this issue, one can use the large horizon distance of the next generation GW detectors\footnote{See~\cite{Mukhopadhyay:2023niv} and references therein for details on the GW and neutrino detectors assumed for the analysis and the results presented in this section.} like Cosmic Explorer (CE), Einstein Telescope (ET), and a combination of the two (CE+ET) to perform \emph{triggered-stacking searches} for high-energy neutrinos from the magnetar remnants of BNS mergers. Triggered stacking searches are prevalent search techniques in the multi-messenger community. In particular GW triggered searches have been explored~\citep{Bartos:2018jco,IceCube:2020xks,IceCube:2022mma} and implemented, for example, the LLAMA pipeline~\citep{Countryman:2019pqq}. 

However, the next generation GW detectors present a challenge as a result of the very large ($z \sim 10 (3.5)$ for CE (ET),~\citealt{galaxies10040090}) horizon distance for detecting BNS mergers. Given the rate of BNS mergers (say $R(z=0) = 300\ {\rm Gpc}^{-3} {\rm yr}^{-1}$) and the redshift dependent rate evolution, it can be concluded that the CE and ET would detect $\mathcal{O}(100)$ events a day. It is evident that this leads to a problem with performing triggered stacking searches due to the large number of triggers. A possible way to solve this issue and collect meaningful triggers was discussed in~\cite{Mukhopadhyay:2023niv,Mukhopadhyay:2024lwq}. The idea relies on the good localization capabilities of the next-generation GW detectors. In this sub-section, we will discuss the prospects of performing GW triggered high-energy neutrino searches at IceCube-Gen2 using CE and ET, for our magnetar model.

The probability to detect more than one neutrino associated with a GW signal can be given by
\begin{align}
q\big( d^{\rm UL}_{\rm GW}, T_{\rm op} \big) &= 1 - {\rm exp}\bigg( -T_{\rm op} I \big( d_{\rm GW}^{\rm UL} \big) \bigg)\,\nonumber\\
I \big( d_{\rm GW}^{\rm UL} \big) = 4\pi &\int_{0}^{d_{\rm GW}^{\rm UL}} d (d_{\rm com}) \frac{T_{\rm op}}{\big( 1+z \big)} R\big( z \big) d_{\rm com}^2 P_{n \geq 1} \big( d_L \big)\,,
\end{align}
where $d^{\rm UL}_{\rm GW} = {\max}\left[{\rm min} \big(d_{\rm GW}^{\rm lim}, d^{\rm hor}_{\rm GW} \big), d^{\rm NG}_{\rm MM}\right]$ is the upper limit from the GW detector from which meaningful triggers can be collected, $d_{\rm GW}^{\rm lim}$ is the limiting distance from the GW detectors based on the localization capability given a choice of the threshold fraction of total sky-area covered $f_{\rm th}$, $d^{\rm hor}_{\rm GW}$ is the horizon distance which we fix at $z \sim 3.5$, $d^{\rm NG}_{\rm MM}$ is the multi-messenger (EM+GW) horizon distance for the next generation detectors, $T_{\rm op}$ is the operation time of the GW detector in the observer frame (on Earth), and the integral is performed over the comoving distance ($d_{\rm com}$). In the next generation of multi-messenger detectors, optical detectors such as the Vera C. Rubin Observatory's Legacy Survey of Space and Time (LSST)~\citep{LSST:2008ijt} can detect kilonovae at least upto $\sim 500$ Mpc. The infrared telescope Nancy Grace Roman Space Telescope~\citep{Foley:2019dck} with a kilonova detection horizon of $\sim 1$ Gpc~\citep{DES:2017dgt} also has very good detection prospects. Besides the improved second generation of GW detectors will have distance horizons up to $\sim 500$ Mpc. As a result of these, we can have better localization for nearby sources. To take this effect into account, we fix $d^{\rm NG}_{\rm MM} = 500$ Mpc.

The declination integrated probability to detect at least one neutrino as a function of luminosity distance is given by $P_{n \geq 1} (d_L)$. The number of $\nu_\mu$ and $\bar{\nu}_\mu$ detected is given by Equation~\eqref{eq:nmunu} and the flux of the same $\phi_{\nu_\mu + \bar{\nu}_\mu}$ is given by Equation~\eqref{eq:flux} which depends on the luminosity distance and hence has a redshift correction of $(1+z)$. We take into account both tracks~\citep{IceCube:2019cia} and shower~\citep{IceCube:2021mbf} effective areas for point source objects. Although showers have poor angular resolution, in this work we are only interested in the number of events and do not use the localization information from the neutrino detectors. 

The redshift dependent rate of BNS mergers is given by $R(z)$ and we choose the fiducial rate to be $R(z=0) = 300\ {\rm Gpc}^{-3} {\rm yr}^{-1}$~\citep{KAGRA:2021duu}. One of the main ingredients for performing triggered-stacking searches is the size of the time window from the time of trigger. We saw from Figure~\ref{fig:nu} that the neutrino emissions peak at around $\sim 10^{5.5}$ s and $\sim 10^{6}$ s for the fiducial and optimistic cases respectively. Thus, we choose these as the size of our windows for the fiducial and optimistic cases. The large time windows are not ideal for triggered searches since this lowers $d_{\rm GW}^{\rm UL}$ for the GW detectors.

Our results for the prospects of performing triggered searches are shown in Figure~\ref{fig:nu_gw}. The \emph{left panel} highlights the best possible scenario when $f_{\rm mag} = 1$ while the \emph{right panel} shows a realistic and conservative scenario when $f_{\rm mag}^{\rm fid} = 0.1$ and $f_{\rm mag}^{\rm opt} = 0.3$ for the fiducial and optimistic cases respectively. The former is highly unlikely but provides a perspective regarding the upper limits given $f_{\rm mag}$ is highly uncertain. We show the results for the fiducial case using CE+ET assuming a $f_{\rm th} = 10^{-2}$ (dashed orange line). The limiting distance in this case is given by $d_{\rm GW}^{\rm UL} \sim 1448$ Mpc. For a realistic choice of $f_{\rm mag}^{\rm fid}$, it is evident that for the fiducial case, with the best possible combination of GW detectors we have $\sim 50$\% probability of a neutrino detection in an operation time of $\sim 30$ years, but constraining the model parameters in case of non-detection, within reasonable timescales of operation, is not viable in this scenario. For $f_{\rm mag} = 1$ the detection probability jumps to $\sim 99$\% in $\lesssim 30$ years and constraints can be placed at the $3\sigma$ level within similar timescales of joint operation.

The optimistic case (shown in dot-dashed pink) shows that neutrino detection is highly probable within $T_{\rm op} \lesssim 15$ years for $f_{\rm mag} = 1$ while for $f_{\rm mag}^{\rm opt}$ there is a $\sim 95$\% probability of neutrino detection in $\sim 20$ years. For the latter, a non-detection can provide $2 \sigma$ constraints within similar timescales. Choosing a more stringent limit on the threshold for fraction of sky area covered ($f_{\rm th}$) with a distance limit $d_{\rm GW}^{\rm UL} \sim 500$ Mpc, might help in placing $2 \sigma$ constraints in a timescale of more than $30$ years. Since $f_{\rm th}$ is low, this scenario has the potential for collecting good quality data and has a $\sim 90$\% probability to detect a neutrino within $30-40$ years. This is because a more stringent limit on $f_{\rm th}$ ensures sufficiently low background with respect to triggered searches. Finally it is important to note that given the long operation times, the choice of $f_{\rm th}$ is constrained by backgrounds (see Figure 4 in~\citealt{Mukhopadhyay:2023niv}). However, recall that the neutrino fluences for these sources are peaked for $E_\nu \sim 10^7 - 10^8$ GeV and at these energies the contribution from atmospheric backgrounds are low. Hence for typical GW detector operation times of $\sim 30$ years, $f_{\rm th} \gtrsim 10^{-2}$ can be used~\citep{Mukhopadhyay:2024lwq}.

\section{Summary}
\label{sec:summary}
In this work, we compute the high-energy neutrino signatures from a long-lived magnetar remnant of a BNS merger. We also investigate the prospects of GW triggered high-energy neutrino searches from such sources at the next-generation neutrino detectors like IceCube-Gen2 and GRAND with triggers from CE and ET.

A BNS merger can leave a stable magnetar as its remnant. The resulting ejecta from the merger surrounds the remnant. The magnetar has $\sim$ millisecond spin period and eventually spins down. The spindown energy from the magnetar acts as an energy source. The magnetar wind consists of $e^{+}-e^{-}$ pairs and the dissipation of its kinetic energy by the TS leads to the formation of a nebula. The nebula consists of large fractions of non-thermal photons. The magnetar also has a high dipolar magnetic field ideal for accelerating the ions (or CR protons) extracted from its surface to ultra-high energies. The accelerated CR protons interact with the nebular photons leading to high-energy neutrino production. A schematic of our model is shown in Figure~\ref{fig:model}.

In this work, we modeled the dynamics of such a system and consistently computed the time-evolution of the non-thermal, thermal, and magnetic energies (see Figure~\ref{fig:energies} for details). The relevant timescale associated with the energies are the spindown time and the characteristic timescales associated with photon diffusion from the nebula to the ejecta and from the ejecta. We considered two parameter sets: the fiducial case where the parameter choices are moderate and and the optimistic parameter set which gives stronger neutrino signatures. The details of the parameters are provided in Table~\ref{tab:params}. 

In our model, the CR protons are extracted from the neutron star crust into the magnetosphere, which can be estimated by the Goldreich-Julian number density. The CR protons are then accelerated at the polar cap region with a $\delta$-function injection spectra. A fraction of the CR protons can also be accelerated at the termination shock (TS) region, where the injection spectra is close to $-2$. The former acceleration results due to the potential gap formed in the polar cap region where charge particles can be accelerated. The latter acceleration is a resultant of plasma instabilities and turbulence that can be prevalent in the TS region. However, we concluded that for our choice of parameters, the acceleration at the TS shock is insignificant. In Figures~\ref{fig:timescales} and~\ref{fig:picooling} we show the details about the CR protons, in particular, the cooling and acceleration timescales relevant for neutrino production, the regions where pion cooling could be important, and the time evolution of the CR proton energies in different regions.

We discuss the observational aspects associated with high-energy neutrinos from our model in Section~\ref{sec:res} and Figure~\ref{fig:nu}. The peak neutrino fluence is $\sim 1 \times 10^{-2}\ {\rm GeV cm}^{-2}$ ($\sim 4 \times 10^{-2}\ {\rm GeV cm}^{-2}$) and the peak occurs between $E_\nu \sim 10^7 - 10^8$ GeV for the fiducial (optimistic) scenarios. Most importantly, the neutrino fluence peak occurs roughly after $10^{5.5}$ s ($10^{6}$ s) post the BNS merger in the fiducial (optimistic) scenarios. Assuming the current point source (PS) limit and scaling it to IceCube-Gen2 or assuming the GRAND-200k PS sensitivity limits show that individual neutrino detections from the magnetar sources is unlikely. The diffuse fluxes resulting from our model assuming a conservative fraction for stable magnetar forming remnants from BNS mergers $f_{\rm mag}^{\rm fid} = 0.1$ and $f_{\rm mag}^{\rm opt} = 0.3$ for the fiducial and optimistic cases, are shown in Figure~\ref{fig:diffuse_nu}. The projected 10-year sensitivities for IceCube-Gen2 (optical and radio) and GRAND-200k are comparable to the diffuse fluxes for the optimistic case. However the fiducial case demands a large value of $f_{\rm mag}^{\rm fid}$ to achieve comparable sensitivities. As a complimentary approach, one can also search for neutrinos from such sources using the technique of triggered-stacking searches. In particular, we investigated the prospects of such searches in the light of the next generation of GW detectors -- the Einstein Telescope (ET) and the Cosmic Explorer (CE).

Assuming realistic values of $f_{\rm mag}^{\rm fid}$ and $f_{\rm mag}^{\rm opt}$, we note from Figure~\ref{fig:nu_gw} that for the fiducial parameter using GW triggered-stacking searches, the probability for a coincident neutrino detection is $\sim 50$\% using the combination of CE+ET in $30$ years of joint operation time. However, in case of non-detection no significant constraints can be placed on the parameter space within reasonable timescales. The optimistic scenario provides better prospects, where using $f_{\rm th} = 10^{-4}$, the neutrino detection probability is $\sim 95$\% in a joint operation timescale of $\sim 20$ years and $2\sigma$ level constraints can be placed in case of non-detection within similar timescales. For the most-optimistic case, that is, when $f_{\rm mag} = 1$, the probability of a coincident neutrino detection is $\sim 99.7$\% within 30 years of joint operation and constraints at the $3\sigma$ level can be placed within similar timescales using CE+ET with $f_{\rm th} = 10^{-2}$.
\section{Discussion and Implications}
\label{sec:disc}
In this section we discuss the shortcomings of our work and its implications in the broader sense. We attempted to answer the questions we began with in Section~\ref{sec:intro}. We modeled the magnetar-wind-nebula-ejecta system and computed the plausible neutrino fluences from such a system, along with contributions to the diffuse flux. We also explored GW-triggered search techniques which can potentially result in detecting neutrinos from such magnetar powered sources. This contributes to the broader understanding of neutrino emission and helps in planning detection strategies from these class of sources.

There are several limitations associated with the modeling of the system, as a result of which we make some assumptions and approximations. First, the expression we use for spindown luminosity (see Equation~\ref{eq:lsd}) assumes an alignment between the rotation and magnetic axes of the magnetar. In reality, the two axes will probably be unaligned leading to $\alpha \neq 1$ and introduce some uncertainty to the spindown luminosity and energy. However, this will change our results at most by a factor of a few. Next, we assume that a constant fraction of potential gap is used for proton acceleration. It is well-known that plasma physics plays a very important role in the magnetosphere of the magnetar. The fraction of the potential gap experienced by the CR protons at the polar cap region ($\eta_{\rm gap}$) could be variable. Depending on this the maximum energy of the accelerated protons in the polar cap region could vary (see Equation~\ref{eq:epcmax}). Next, we discuss the uncertainties associated with the modeling of the effective radius $R$. We use the same value of $R$ for the nebula and the ejecta. Although this approximation is reasonable as long as the thickness of the ejecta is thin, it is unclear whether this condition is satisfied or not. Dedicated multi-dimensional hydrodynamic simulations are necessary to access the validity of this assumption.

We assume the magnetic field in the nebula ($B_{\rm neb}^\prime$) to be turbulent and that the magnetic field strength in the nebula is uniform. In reality the nebular magnetic field strength depends on the distance from the TS and the turbulent magnetic field might decay with distance. Assuming aligned fields are still dominant, the magnetic field strength associated with the toroidal component can depend on the distance from the TS. Furthermore, the magnetic field dissipation parameter ($\epsilon_B$) is highly uncertain and can lead to a higher or lower magnetic energy in the nebula. Next there are also some uncertainties associated with the injection efficiency of the CR protons. The Goldreich-Julian limit we use for our work can breakdown in certain limits of extreme magnetic fields~\citep{Sobyanin:2016acr}. Moreover, fallback matter might enhance the baryon loading factor at the initial stages which might alter the neutrino fluences.

A more accurate calculation can be performed by consistently taking into account the time-dependent pair multiplicity factor $Y$ in the nebula. We fix this to $Y = 0.1$ and assume a saturated state, in accordance with previous literature. The pair multiplicity can alter the photon diffusion times from the nebula to the ejecta, thus introducing some uncertainty in the neutrino fluence. Furthermore, we choose the acceleration efficiency at the TS region $\eta_{\rm acc} = 1$. The efficiency can vary between $1 - 10$, and hence can introduce a factor of a few correction to the cut-off energies at the TS region. However, this is insignificant for our work since most the protons are accelerated in the polar cap region.

The composition of the surface (crust) of a neutron star is not very well understood~\citep{Ozel:2012wu}. It is plausible that after the merger the neutron star crust is mainly composed of heavier nuclei like iron. However, in this work, we do not consider acceleration of iron-like nuclei or any heavy nuclei (see~\citealt{Blasi:2000xm} where acceleration of iron nuclei was considered). We strictly only consider protons ($Z=1$) extracted from the crust for acceleration and subsequent neutrino production. Various effects like photodissociation and photopion production cross-section would need to be carefully accounted for in the case of heavier nuclei. Usually, the photopion cross-section per nucleon is similar to that for protons, and we expect comparable neutrino fluences unless photodissociation completely alters the maximum energy. Furthermore, although a higher $Z$ increases the energy of CR nuclei, the resulting neutrino energy is determined by the energy per nucleon, that is the CR Lorentz factor, which is similar to that for protons. Since we determine the injected CR number using the GJ charge density, the number of protons are similar in either case. Thus, based on our qualitative understanding, the effect of heavy nuclei on the neutrino fluence should be moderate as long as photodissociation completely destroys the nuclei and lowers the Lorentz factor. However, the variations in the neutrino fluence due to heavier nuclei is non-trivial and we leave that for future work.

For the GW-triggered searches, it is important to note that the improved second generation of GW detectors along with the upcoming EM telescopes can help in reducing the uncertainty region associated with the localization of the merger. In particular, EM telescopes can greatly narrow down the source location. The BNS merger will be accompanied by emissions in the radio, optical, X-ray, and gamma ray bands. For radio, X-ray, and gamma-ray counterparts, GRB prompt emission and afterglows are strong, but these emissions are beamed. Thus, we will only have a fraction of BNS mergers from which such signals can be detected. Therefore, optical counterparts as a result of the mergers which could be detected at Vera C. Rubin Observatory's Legacy Survey of Space and Time (LSST), would provide the best prospects to localize the nearby BNS mergers events.

Our work besides introducing a novel treatment to studying high-energy neutrino emissions from magnetar remnant of BNS mergers, also largely improves on previous work by~\cite{Fang:2017tla}. We consistently take into account the relativistic effects in solving for the dynamics of the magnetar-nebula-ejecta system. Moreover, we consistently solve for $\mathcal{A}$ assuming a composition of the kilonova ejecta, while the previous work assumed $\mathcal{A} = 0$. This modifies the results significantly since choosing $\mathcal{A} = 0$ assumes perfectly absorbing boundaries. We consider CR proton accelerations at the polar cap region with a $\delta-$function injection spectra. In~\cite{Fang:2017tla} the authors considered a single power-law injection with index of $-1$ and acceleration at the polar cap region, which is realized at late times when the magnetar has significantly spun down. For our time intervals of interest, the magnetar spin frequency is not drastically decreased, and hence a $\delta-$function injection spectra (see Equation~\ref{eq:crspec_inj}) is more appropriate. We also improve on the previous work by solving the CR transport equation in the steady state assumption. Moreover, we incorporate a proton energy-dependent CR diffusion coefficient (see Equation~\ref{eq:crdiffusion}) unlike the previous work where the authors fixed the diffusion timescale to $R/c$. Finally the fiducial choice of parameters assumed in~\cite{Fang:2017tla} is optimistic leading to neutrino fluences which are $\sim 1$ order of magnitude larger than what we have for our optimistic case. In fact, the diffuse neutrino flux results for the fiducial parameter choices in the previous work has been excluded by the current IceCube diffuse flux limits~\citep{IceCube:2018fhm,IceCube:2023luu}; however the conservative models are still consistent with the current data.

Our work highlights that stable long-lived magnetar remnants resulting from BNS mergers may not be ideal sources  of high-energy neutrino signatures. However, such objects may exist and hence understanding the relevant particle acceleration and neutrino production sites is of crucial importance. Besides the next generation of multi-messenger astronomy enabled by neutrino detectors like IceCube-Gen2, KM3NeT and GW detectors like CE and ET make exploring the prospects of multi-messenger detections from these source classes very timely. In the future detecting neutrino signatures from such objects can give vast insights at the micro-physics level, on the behaviour of plasma in highly magnetized environment, on the dynamics of BNS mergers including the EOS and kilonovae properties. The plethora of information that could be conveyed as a result of neutrino detections from these magnetar remnants of BNS mergers serves as a motivation for modeling such sources. Our work provides some key insights on high-energy neutrino signatures that could be expected and sheds light on the possibilities of synergic GW neutrino observations in the next era of multimessenger astrophysics.
%
%
\begin{acknowledgements}
We thank Mukul Bhattacharya, Ke Fang, Rossella Gamba, Kazumi Kashiyama, Shota Kisaka, Kohta Murase, Alexander Philippov, and David Radice for useful discussions. We are particularly grateful to Ke Fang and the anonymous referee for their insightful comments that lead to an improved version of this manuscript. M.\,M. wishes to thank the Astronomical Institute at Tohoku University, for their hospitality where a major part of this work was completed.
M.\,M. is supported by NSF Grant No. AST-2108466. M.\,M. also acknowledges support from the 
Institute for Gravitation and the Cosmos (IGC) Postdoctoral Fellowship. S.S.K. acknowledges the support by KAKENHI No. 22K14028, No. 21H04487, No. 23H04899, and the Tohoku Initiative for Fostering Global Researchers for Interdisciplinary Sciences (TI-FRIS) of MEXT’s Strategic Professional Development Program for Young Researchers. B.D.M. acknowledges support from the NASA Astrophysics Theory Program (grant number AST-2002577) and NASA Fermi Guest Investigator Program (grant number 80NSSC22K1574). The Flatiron Institute is supported by the Simons Foundation.
\end{acknowledgements}
\bibliographystyle{aasjournal}
\bibliography{refs}

\begin{thebibliography}{}
\expandafter\ifx\csname natexlab\endcsname\relax\def\natexlab#1{#1}\fi
\providecommand{\url}[1]{\href{#1}{#1}}
\providecommand{\dodoi}[1]{doi:~\href{http://doi.org/#1}{\nolinkurl{#1}}}
\providecommand{\doeprint}[1]{\href{http://ascl.net/#1}{\nolinkurl{http://ascl.net/#1}}}
\providecommand{\doarXiv}[1]{\href{https://arxiv.org/abs/#1}{\nolinkurl{https://arxiv.org/abs/#1}}}

\bibitem[{Aartsen {et~al.}(2018)}]{IceCube:2018fhm}
Aartsen, M.~G., {et~al.} 2018, Phys. Rev. D, 98, 062003, \dodoi{10.1103/PhysRevD.98.062003}

\bibitem[{Aartsen {et~al.}(2020{\natexlab{a}})}]{IceCube:2020xks}
---. 2020{\natexlab{a}}, Astrophys. J. Lett., 898, L10, \dodoi{10.3847/2041-8213/ab9d24}

\bibitem[{Aartsen {et~al.}(2020{\natexlab{b}})}]{IceCube:2019cia}
---. 2020{\natexlab{b}}, Phys. Rev. Lett., 124, 051103, \dodoi{10.1103/PhysRevLett.124.051103}

\bibitem[{Aartsen {et~al.}(2021)}]{IceCube-Gen2:2020qha}
---. 2021, J. Phys. G, 48, 060501, \dodoi{10.1088/1361-6471/abbd48}

\bibitem[{Abbasi {et~al.}(2023{\natexlab{a}})}]{IceCube:2022mma}
Abbasi, R., {et~al.} 2023{\natexlab{a}}, Astrophys. J., 944, 80, \dodoi{10.3847/1538-4357/aca5fc}

\bibitem[{Abbasi {et~al.}(2023{\natexlab{b}})}]{IceCube:2023luu}
---. 2023{\natexlab{b}}, PoS, ICRC2023, 1149, \dodoi{10.22323/1.444.1149}

\bibitem[{Abbott {et~al.}(2017{\natexlab{a}})}]{LIGOScientific:2017vwq}
Abbott, B.~P., {et~al.} 2017{\natexlab{a}}, Phys. Rev. Lett., 119, 161101, \dodoi{10.1103/PhysRevLett.119.161101}

\bibitem[{Abbott {et~al.}(2017{\natexlab{b}})}]{LIGOScientific:2017zic}
---. 2017{\natexlab{b}}, Astrophys. J. Lett., 848, L13, \dodoi{10.3847/2041-8213/aa920c}

\bibitem[{Abbott {et~al.}(2017{\natexlab{c}})}]{LIGOScientific:2017ync}
---. 2017{\natexlab{c}}, Astrophys. J. Lett., 848, L12, \dodoi{10.3847/2041-8213/aa91c9}

\bibitem[{Abbott {et~al.}(2023)}]{KAGRA:2021duu}
Abbott, R., {et~al.} 2023, Phys. Rev. X, 13, 011048, \dodoi{10.1103/PhysRevX.13.011048}

\bibitem[{Albert {et~al.}(2017)}]{ANTARES:2017bia}
Albert, A., {et~al.} 2017, Astrophys. J. Lett., 850, L35, \dodoi{10.3847/2041-8213/aa9aed}

\bibitem[{\'Alvarez-Mu\~niz {et~al.}(2020)}]{GRAND:2018iaj}
\'Alvarez-Mu\~niz, J., {et~al.} 2020, Sci. China Phys. Mech. Astron., 63, 219501, \dodoi{10.1007/s11433-018-9385-7}

\bibitem[{Arons(2003)}]{Arons:2002yj}
Arons, J. 2003, Astrophys. J., 589, 871, \dodoi{10.1086/374776}

\bibitem[{Ascenzi {et~al.}(2021)Ascenzi, Oganesyan, Branchesi, \& Ciolfi}]{Ascenzi:2020xqi}
Ascenzi, S., Oganesyan, G., Branchesi, M., \& Ciolfi, R. 2021, J. Plasma Phys., 87, 845870102, \dodoi{10.1017/S0022377820001646}

\bibitem[{Baiotti(2019)}]{Baiotti:2019sew}
Baiotti, L. 2019, Prog. Part. Nucl. Phys., 109, 103714, \dodoi{10.1016/j.ppnp.2019.103714}

\bibitem[{Baiotti \& Rezzolla(2017)}]{Baiotti:2016qnr}
Baiotti, L., \& Rezzolla, L. 2017, Rept. Prog. Phys., 80, 096901, \dodoi{10.1088/1361-6633/aa67bb}

\bibitem[{Bartos {et~al.}(2019)Bartos, Veske, Keivani, Marka, Countryman, Blaufuss, Finley, \& Marka}]{Bartos:2018jco}
Bartos, I., Veske, D., Keivani, A., {et~al.} 2019, Phys. Rev. D, 100, 083017, \dodoi{10.1103/PhysRevD.100.083017}

\bibitem[{Becker(2008)}]{BECKER2008173}
Becker, J.~K. 2008, Physics Reports, 458, 173, \dodoi{https://doi.org/10.1016/j.physrep.2007.10.006}

\bibitem[{Blandford \& Eichler(1987)}]{Blandford:1987pw}
Blandford, R., \& Eichler, D. 1987, Phys. Rept., 154, 1, \dodoi{10.1016/0370-1573(87)90134-7}

\bibitem[{Blandford(2002)}]{Blandford2002}
Blandford, R.~D. 2002, in Lighthouses of the Universe: The Most Luminous Celestial Objects and Their Use for Cosmology, ed. M.~Gilfanov, R.~Sunyeav, \& E.~Churazov (Berlin, Heidelberg: Springer Berlin Heidelberg), 381--404

\bibitem[{Blasi {et~al.}(2000)Blasi, Epstein, \& Olinto}]{Blasi:2000xm}
Blasi, P., Epstein, R.~I., \& Olinto, A.~V. 2000, Astrophys. J. Lett., 533, L123, \dodoi{10.1086/312626}

\bibitem[{Blinnikov {et~al.}(2018)Blinnikov, Novikov, Perevodchikova, \& Polnarev}]{Blinnikov:2018boq}
Blinnikov, S.~I., Novikov, I.~D., Perevodchikova, T.~V., \& Polnarev, A.~G. 2018.
\newblock \doarXiv{1808.05287}

\bibitem[{{Bogovalov} {et~al.}(2005){Bogovalov}, {Chechetkin}, {Koldoba}, \& {Ustyugova}}]{Bogovalov2005}
{Bogovalov}, S.~V., {Chechetkin}, V.~M., {Koldoba}, A.~V., \& {Ustyugova}, G.~V. 2005, \mnras, 358, 705, \dodoi{10.1111/j.1365-2966.2004.08592.x}

\bibitem[{Bonazzola \& Gourgoulhon(1996)}]{Bonazzola:1995rb}
Bonazzola, S., \& Gourgoulhon, E. 1996, Astron. Astrophys., 312, 675.
\newblock \doarXiv{astro-ph/9602107}

\bibitem[{Carpio {et~al.}(2020)Carpio, Murase, Reno, Sarcevic, \& Stasto}]{Carpio:2020wzg}
Carpio, J.~A., Murase, K., Reno, M.~H., Sarcevic, I., \& Stasto, A. 2020, Phys. Rev. D, 102, 103001, \dodoi{10.1103/PhysRevD.102.103001}

\bibitem[{Cerutti \& Beloborodov(2017)}]{Cerutti:2016ttn}
Cerutti, B., \& Beloborodov, A. 2017, Space Sci. Rev., 207, 111, \dodoi{10.1007/s11214-016-0315-7}

\bibitem[{Cerutti {et~al.}(2015)Cerutti, Philippov, Parfrey, \& Spitkovsky}]{Cerutti:2014ysa}
Cerutti, B., Philippov, A., Parfrey, K., \& Spitkovsky, A. 2015, Mon. Not. Roy. Astron. Soc., 448, 606, \dodoi{10.1093/mnras/stv042}

\bibitem[{Chen \& Beloborodov(2014)}]{Chen:2014dva}
Chen, A.~Y., \& Beloborodov, A.~M. 2014, Astrophys. J. Lett., 795, L22, \dodoi{10.1088/2041-8205/795/1/L22}

\bibitem[{{Chodorowski} {et~al.}(1992){Chodorowski}, {Zdziarski}, \& {Sikora}}]{1992ApJ...400..181C}
{Chodorowski}, M.~J., {Zdziarski}, A.~A., \& {Sikora}, M. 1992, \apj, 400, 181, \dodoi{10.1086/171984}

\bibitem[{Chornock {et~al.}(2017)}]{Chornock:2017sdf}
Chornock, R., {et~al.} 2017, Astrophys. J. Lett., 848, L19, \dodoi{10.3847/2041-8213/aa905c}

\bibitem[{Combi \& Siegel(2023)}]{Combi:2023yav}
Combi, L., \& Siegel, D.~M. 2023, Phys. Rev. Lett., 131, 231402, \dodoi{10.1103/PhysRevLett.131.231402}

\bibitem[{Coulter {et~al.}(2017)}]{Coulter:2017wya}
Coulter, D.~A., {et~al.} 2017, Science, 358, 1556, \dodoi{10.1126/science.aap9811}

\bibitem[{Countryman {et~al.}(2019)Countryman, Keivani, Bartos, Marka, Kintscher, Corley, Blaufuss, Finley, \& Marka}]{Countryman:2019pqq}
Countryman, S., Keivani, A., Bartos, I., {et~al.} 2019.
\newblock \doarXiv{1901.05486}

\bibitem[{Dall'Osso {et~al.}(2009)Dall'Osso, Shore, \& Stella}]{DallOsso:2008kll}
Dall'Osso, S., Shore, S.~N., \& Stella, L. 2009, Mon. Not. Roy. Astron. Soc., 398, 1869, \dodoi{10.1111/j.1365-2966.2008.14054.x}

\bibitem[{D'Avanzo {et~al.}(2018)}]{DAvanzo:2018zyz}
D'Avanzo, P., {et~al.} 2018, Astron. Astrophys., 613, L1, \dodoi{10.1051/0004-6361/201832664}

\bibitem[{Del~Zanna {et~al.}(2004)Del~Zanna, Amato, \& Bucciantini}]{DelZanna:2004aq}
Del~Zanna, L., Amato, E., \& Bucciantini, N. 2004, Astron. Astrophys., 421, 1063, \dodoi{10.1051/0004-6361:20035936}

\bibitem[{{Dermer} \& {Menon}(2009)}]{DermerMenon}
{Dermer}, C.~D., \& {Menon}, G. 2009, {High Energy Radiation from Black Holes: Gamma Rays, Cosmic Rays, and Neutrinos}

\bibitem[{Dietrich {et~al.}(2021)Dietrich, Hinderer, \& Samajdar}]{Dietrich:2020eud}
Dietrich, T., Hinderer, T., \& Samajdar, A. 2021, Gen. Rel. Grav., 53, 27, \dodoi{10.1007/s10714-020-02751-6}

\bibitem[{Drout {et~al.}(2017)}]{Drout:2017ijr}
Drout, M.~R., {et~al.} 2017, Science, 358, 1570, \dodoi{10.1126/science.aaq0049}

\bibitem[{{Eichler} {et~al.}(1989){Eichler}, {Livio}, {Piran}, \& {Schramm}}]{1989Natur.340..126E}
{Eichler}, D., {Livio}, M., {Piran}, T., \& {Schramm}, D.~N. 1989, \nat, 340, 126, \dodoi{10.1038/340126a0}

\bibitem[{Fang {et~al.}(2014)Fang, Kotera, Murase, \& Olinto}]{Fang:2013vla}
Fang, K., Kotera, K., Murase, K., \& Olinto, A.~V. 2014, Phys. Rev. D, 90, 103005, \dodoi{10.1103/PhysRevD.90.103005}

\bibitem[{Fang \& Metzger(2017)}]{Fang:2017tla}
Fang, K., \& Metzger, B.~D. 2017, Astrophys. J., 849, 153, \dodoi{10.3847/1538-4357/aa8b6a}

\bibitem[{Fang {et~al.}(2019)Fang, Metzger, Murase, Bartos, \& Kotera}]{Fang:2018hjp}
Fang, K., Metzger, B.~D., Murase, K., Bartos, I., \& Kotera, K. 2019, Astrophys. J., 878, 34, \dodoi{10.3847/1538-4357/ab1b72}

\bibitem[{Fern\'andez \& Metzger(2016)}]{Fernandez:2015use}
Fern\'andez, R., \& Metzger, B.~D. 2016, Ann. Rev. Nucl. Part. Sci., 66, 23, \dodoi{10.1146/annurev-nucl-102115-044819}

\bibitem[{Foley {et~al.}(2019)}]{Foley:2019dck}
Foley, R.~J., {et~al.} 2019.
\newblock \doarXiv{1903.04582}

\bibitem[{Fujibayashi {et~al.}(2020)Fujibayashi, Wanajo, Kiuchi, Kyutoku, Sekiguchi, \& Shibata}]{Fujibayashi:2020dvr}
Fujibayashi, S., Wanajo, S., Kiuchi, K., {et~al.} 2020, Astrophys. J., 901, 122, \dodoi{10.3847/1538-4357/abafc2}

\bibitem[{Gao {et~al.}(2013)Gao, Zhang, Wu, \& Dai}]{Gao:2013rxa}
Gao, H., Zhang, B., Wu, X.-F., \& Dai, Z.-G. 2013, Phys. Rev. D, 88, 043010, \dodoi{10.1103/PhysRevD.88.043010}

\bibitem[{Ghirlanda {et~al.}(2019)}]{Ghirlanda:2018uyx}
Ghirlanda, G., {et~al.} 2019, Science, 363, 968, \dodoi{10.1126/science.aau8815}

\bibitem[{Goldreich \& Julian(1969)}]{Goldreich:1969sb}
Goldreich, P., \& Julian, W.~H. 1969, Astrophys. J., 157, 869, \dodoi{10.1086/150119}

\bibitem[{Goldstein {et~al.}(2017)}]{Goldstein:2017mmi}
Goldstein, A., {et~al.} 2017, Astrophys. J. Lett., 848, L14, \dodoi{10.3847/2041-8213/aa8f41}

\bibitem[{Gruzinov(1999)}]{Gruzinov:1999aza}
Gruzinov, A. 1999.
\newblock \doarXiv{astro-ph/9902288}

\bibitem[{Gunn \& Ostriker(1969)}]{Gunn:1969ej}
Gunn, J.~E., \& Ostriker, J.~P. 1969, Phys. Rev. Lett., 22, 728, \dodoi{10.1103/PhysRevLett.22.728}

\bibitem[{Guo {et~al.}(2020)Guo, Liu, Li, Li, Daughton, \& Kilian}]{Guo:2020fni}
Guo, F., Liu, Y.-H., Li, X., {et~al.} 2020, Phys. Plasmas, 27, 080501, \dodoi{10.1063/5.0012094}

\bibitem[{Haggard {et~al.}(2017)Haggard, Nynka, Ruan, Kalogera, Bradley~Cenko, Evans, \& Kennea}]{Haggard:2017qne}
Haggard, D., Nynka, M., Ruan, J.~J., {et~al.} 2017, Astrophys. J. Lett., 848, L25, \dodoi{10.3847/2041-8213/aa8ede}

\bibitem[{Hall.(2022)}]{galaxies10040090}
Hall., E.~D. 2022, Galaxies, 10, \dodoi{10.3390/galaxies10040090}

\bibitem[{Hallinan {et~al.}(2017)}]{Hallinan:2017woc}
Hallinan, G., {et~al.} 2017, Science, 358, 1579, \dodoi{10.1126/science.aap9855}

\bibitem[{Harari {et~al.}(2014)Harari, Mollerach, \& Roulet}]{Harari:2013pea}
Harari, D., Mollerach, S., \& Roulet, E. 2014, Phys. Rev. D, 89, 123001, \dodoi{10.1103/PhysRevD.89.123001}

\bibitem[{Hayato {et~al.}(2018)}]{Super-Kamiokande:2018dbf}
Hayato, Y., {et~al.} 2018, Astrophys. J. Lett., 857, L4, \dodoi{10.3847/2041-8213/aabaca}

\bibitem[{Hotokezaka {et~al.}(2016)Hotokezaka, Wanajo, Tanaka, Bamba, Terada, \& Piran}]{Hotokezaka:2015cma}
Hotokezaka, K., Wanajo, S., Tanaka, M., {et~al.} 2016, Mon. Not. Roy. Astron. Soc., 459, 35, \dodoi{10.1093/mnras/stw404}

\bibitem[{{IceCube-Gen2 Collaboration}(2024)}]{Gen2_TDR}
{IceCube-Gen2 Collaboration}. 2024, {IceCube-Gen2 Technical Design Report}, Tech. rep.
\newblock \url{https://icecube-gen2.wisc.edu/science/publications/tdr/}

\bibitem[{Ivezi\'c {et~al.}(2019)}]{LSST:2008ijt}
Ivezi\'c, v., {et~al.} 2019, Astrophys. J., 873, 111, \dodoi{10.3847/1538-4357/ab042c}

\bibitem[{Kasen \& Bildsten(2010)}]{Kasen:2009tg}
Kasen, D., \& Bildsten, L. 2010, Astrophys. J., 717, 245, \dodoi{10.1088/0004-637X/717/1/245}

\bibitem[{Kaspi \& Beloborodov(2017)}]{Kaspi:2017fwg}
Kaspi, V.~M., \& Beloborodov, A. 2017, Ann. Rev. Astron. Astrophys., 55, 261, \dodoi{10.1146/annurev-astro-081915-023329}

\bibitem[{Kelner \& Aharonian(2008)}]{Kelner:2008ke}
Kelner, S.~R., \& Aharonian, F.~A. 2008, Phys. Rev. D, 78, 034013, \dodoi{10.1103/PhysRevD.82.099901}

\bibitem[{Kilpatrick {et~al.}(2017)}]{Kilpatrick:2017mhz}
Kilpatrick, C.~D., {et~al.} 2017, Science, 358, 1583, \dodoi{10.1126/science.aaq0073}

\bibitem[{Kimura(2023)}]{Kimura:2022zyg}
Kimura, S.~S. 2023, 433, \dodoi{10.1142/9789811282645_0009}

\bibitem[{Kimura {et~al.}(2019)Kimura, Murase, \& M\'esz\'aros}]{Kimura:2019yjo}
Kimura, S.~S., Murase, K., \& M\'esz\'aros, P. 2019, Phys. Rev. D, 100, 083014, \dodoi{10.1103/PhysRevD.100.083014}

\bibitem[{Komissarov(2013)}]{Komissarov:2012sr}
Komissarov, S.~S. 2013, Mon. Not. Roy. Astron. Soc., 428, 2459, \dodoi{10.1093/mnras/sts214}

\bibitem[{Komissarov \& Lyubarsky(2003)}]{Komissarov:2003tg}
Komissarov, S.~S., \& Lyubarsky, Y.~E. 2003, Mon. Not. Roy. Astron. Soc., 344, L93, \dodoi{10.1046/j.1365-8711.2003.07097.x}

\bibitem[{Kotera {et~al.}(2015)Kotera, Amato, \& Blasi}]{Kotera:2015pya}
Kotera, K., Amato, E., \& Blasi, P. 2015, JCAP, 08, 026, \dodoi{10.1088/1475-7516/2015/08/026}

\bibitem[{Lemoine {et~al.}(2015)Lemoine, Kotera, \& P\'etri}]{Lemoine:2014ala}
Lemoine, M., Kotera, K., \& P\'etri, J. 2015, JCAP, 07, 016, \dodoi{10.1088/1475-7516/2015/07/016}

\bibitem[{Lipunov {et~al.}(2017)}]{Lipunov:2017dwd}
Lipunov, V.~M., {et~al.} 2017, Astrophys. J. Lett., 850, L1, \dodoi{10.3847/2041-8213/aa92c0}

\bibitem[{Maggiore {et~al.}(2020)}]{Maggiore:2019uih}
Maggiore, M., {et~al.} 2020, JCAP, 03, 050, \dodoi{10.1088/1475-7516/2020/03/050}

\bibitem[{Margalit {et~al.}(2022)Margalit, Jermyn, Metzger, Roberts, \& Quataert}]{Margalit:2022rde}
Margalit, B., Jermyn, A.~S., Metzger, B.~D., Roberts, L.~F., \& Quataert, E. 2022, Astrophys. J., 939, 51, \dodoi{10.3847/1538-4357/ac8b01}

\bibitem[{Margutti {et~al.}(2017)}]{Margutti:2017cjl}
Margutti, R., {et~al.} 2017, Astrophys. J. Lett., 848, L20, \dodoi{10.3847/2041-8213/aa9057}

\bibitem[{{Metzger}(2019)}]{Metzger2019}
{Metzger}, B.~D. 2019, Living Reviews in Relativity, 23, 1, \dodoi{10.1007/s41114-019-0024-0}

\bibitem[{Metzger \& Fern\'andez(2014)}]{Metzger:2014ila}
Metzger, B.~D., \& Fern\'andez, R. 2014, Mon. Not. Roy. Astron. Soc., 441, 3444, \dodoi{10.1093/mnras/stu802}

\bibitem[{Metzger \& Piro(2014)}]{Metzger:2013cha}
Metzger, B.~D., \& Piro, A.~L. 2014, Mon. Not. Roy. Astron. Soc., 439, 3916, \dodoi{10.1093/mnras/stu247}

\bibitem[{Metzger {et~al.}(2014)Metzger, Vurm, Hascoet, \& Beloborodov}]{Metzger:2013kia}
Metzger, B.~D., Vurm, I., Hascoet, R., \& Beloborodov, A.~M. 2014, Mon. Not. Roy. Astron. Soc., 437, 703, \dodoi{10.1093/mnras/stt1922}

\bibitem[{Metzger {et~al.}(2010)Metzger, Martinez-Pinedo, Darbha, Quataert, Arcones, Kasen, Thomas, Nugent, Panov, \& Zinner}]{Metzger:2010sy}
Metzger, B.~D., Martinez-Pinedo, G., Darbha, S., {et~al.} 2010, Mon. Not. Roy. Astron. Soc., 406, 2650, \dodoi{10.1111/j.1365-2966.2010.16864.x}

\bibitem[{{M{\"u}cke} {et~al.}(2000){M{\"u}cke}, {Engel}, {Rachen}, {Protheroe}, \& {Stanev}}]{2000CoPhC.124..290M}
{M{\"u}cke}, A., {Engel}, R., {Rachen}, J.~P., {Protheroe}, R.~J., \& {Stanev}, T. 2000, Computer Physics Communications, 124, 290, \dodoi{10.1016/S0010-4655(99)00446-4}

\bibitem[{Mukhopadhyay \& Kimura(2025)}]{Mukhopadhyay:2025tvz}
Mukhopadhyay, M., \& Kimura, S.~S. 2025, Astrophys. J. Lett., 989, L41, \dodoi{10.3847/2041-8213/adf285}

\bibitem[{Mukhopadhyay {et~al.}(2024{\natexlab{a}})Mukhopadhyay, Kimura, \& Murase}]{Mukhopadhyay:2023niv}
Mukhopadhyay, M., Kimura, S.~S., \& Murase, K. 2024{\natexlab{a}}, Phys. Rev. D, 109, 043053, \dodoi{10.1103/PhysRevD.109.043053}

\bibitem[{Mukhopadhyay {et~al.}(2024{\natexlab{b}})Mukhopadhyay, Kotera, Wissel, Murase, \& Kimura}]{Mukhopadhyay:2024lwq}
Mukhopadhyay, M., Kotera, K., Wissel, S., Murase, K., \& Kimura, S.~S. 2024{\natexlab{b}}.
\newblock \doarXiv{2406.19440}

\bibitem[{Murase \& Bartos(2019)}]{Murase:2019tjj}
Murase, K., \& Bartos, I. 2019, Ann. Rev. Nucl. Part. Sci., 69, 477, \dodoi{10.1146/annurev-nucl-101918-023510}

\bibitem[{Murase {et~al.}(2015)Murase, Kashiyama, Kiuchi, \& Bartos}]{Murase:2014bfa}
Murase, K., Kashiyama, K., Kiuchi, K., \& Bartos, I. 2015, Astrophys. J., 805, 82, \dodoi{10.1088/0004-637X/805/1/82}

\bibitem[{Murase {et~al.}(2009)Murase, Meszaros, \& Zhang}]{Murase:2009pg}
Murase, K., Meszaros, P., \& Zhang, B. 2009, Phys. Rev. D, 79, 103001, \dodoi{10.1103/PhysRevD.79.103001}

\bibitem[{Murase \& Nagataki(2006)}]{Murase:2006dr}
Murase, K., \& Nagataki, S. 2006, Phys. Rev. Lett., 97, 051101, \dodoi{10.1103/PhysRevLett.97.051101}

\bibitem[{{Murase} {et~al.}(2021){Murase}, {Omand}, {Coppejans}, {Nagai}, {Bower}, {Chornock}, {Fox}, {Kashiyama}, {Law}, {Margutti}, \& {M{\'e}sz{\'a}ros}}]{2021MNRAS.508...44M}
{Murase}, K., {Omand}, C. M.~B., {Coppejans}, D.~L., {et~al.} 2021, \mnras, 508, 44, \dodoi{10.1093/mnras/stab2506}

\bibitem[{Nakar(2020)}]{Nakar:2019fza}
Nakar, E. 2020, Phys. Rept., 886, 1, \dodoi{10.1016/j.physrep.2020.08.008}

\bibitem[{Nicholl {et~al.}(2017)Nicholl, Berger, Margutti, Blanchard, Guillochon, Leja, \& Chornock}]{Nicholl:2017mnb}
Nicholl, M., Berger, E., Margutti, R., {et~al.} 2017, Astrophys. J. Lett., 845, L8, \dodoi{10.3847/2041-8213/aa82b1}

\bibitem[{Nicholl {et~al.}(2018)}]{Nicholl:2018cam}
Nicholl, M., {et~al.} 2018, Astrophys. J. Lett., 866, L24, \dodoi{10.3847/2041-8213/aae70d}

\bibitem[{{Ostriker} \& {Gunn}(1969)}]{OstrikerGunn1969}
{Ostriker}, J.~P., \& {Gunn}, J.~E. 1969, \apj, 157, 1395, \dodoi{10.1086/150160}

\bibitem[{Owen {et~al.}(1998)Owen, Lindblom, Cutler, Schutz, Vecchio, \& Andersson}]{Owen:1998xg}
Owen, B.~J., Lindblom, L., Cutler, C., {et~al.} 1998, Phys. Rev. D, 58, 084020, \dodoi{10.1103/PhysRevD.58.084020}

\bibitem[{Ozel(2013)}]{Ozel:2012wu}
Ozel, F. 2013, Rept. Prog. Phys., 76, 016901, \dodoi{10.1088/0034-4885/76/1/016901}

\bibitem[{Philippov \& Spitkovsky(2014)}]{Philippov:2013tpa}
Philippov, A.~A., \& Spitkovsky, A. 2014, Astrophys. J. Lett., 785, L33, \dodoi{10.1088/2041-8205/785/2/L33}

\bibitem[{Pian {et~al.}(2017)}]{Pian:2017gtc}
Pian, E., {et~al.} 2017, Nature, 551, 67, \dodoi{10.1038/nature24298}

\bibitem[{Piro \& Kollmeier(2016)}]{Piro:2016jaq}
Piro, A.~L., \& Kollmeier, J.~A. 2016, Astrophys. J., 826, 97, \dodoi{10.3847/0004-637X/826/1/97}

\bibitem[{Quirola-V\'asquez {et~al.}(2023)}]{Quirola-Vasquez:2023eye}
Quirola-V\'asquez, J., {et~al.} 2023, Astron. Astrophys., 675, A44, \dodoi{10.1051/0004-6361/202345912}

\bibitem[{Radice {et~al.}(2020)Radice, Bernuzzi, \& Perego}]{Radice:2020ddv}
Radice, D., Bernuzzi, S., \& Perego, A. 2020, Ann. Rev. Nucl. Part. Sci., 70, 95, \dodoi{10.1146/annurev-nucl-013120-114541}

\bibitem[{Reitze {et~al.}(2019)}]{Reitze:2019iox}
Reitze, D., {et~al.} 2019, Bull. Am. Astron. Soc., 51, 035, \dodoi{10.48550/arXiv.1907.04833}

\bibitem[{Rosswog(2015)}]{Rosswog:2015nja}
Rosswog, S. 2015, Int. J. Mod. Phys. D, 24, 1530012, \dodoi{10.1142/S0218271815300128}

\bibitem[{Ruiz {et~al.}(2021)Ruiz, Shapiro, \& Tsokaros}]{Ruiz:2021gsv}
Ruiz, M., Shapiro, S.~L., \& Tsokaros, A. 2021, Front. Astron. Space Sci., 8, 39, \dodoi{10.3389/fspas.2021.656907}

\bibitem[{Sarin \& Lasky(2021)}]{Sarin:2020gxb}
Sarin, N., \& Lasky, P.~D. 2021, Gen. Rel. Grav., 53, 59, \dodoi{10.1007/s10714-021-02831-1}

\bibitem[{Savchenko {et~al.}(2017)}]{Savchenko:2017ffs}
Savchenko, V., {et~al.} 2017, Astrophys. J. Lett., 848, L15, \dodoi{10.3847/2041-8213/aa8f94}

\bibitem[{Sclafani {et~al.}(2021)}]{IceCube:2021mbf}
Sclafani, S., {et~al.} 2021, PoS, ICRC2021, 1150, \dodoi{10.22323/1.395.1150}

\bibitem[{Scolnic {et~al.}(2018)}]{DES:2017dgt}
Scolnic, D., {et~al.} 2018, Astrophys. J. Lett., 852, L3, \dodoi{10.3847/2041-8213/aa9d82}

\bibitem[{{Shapiro} \& {Teukolsky}(1983)}]{1983bhwd.book.....S}
{Shapiro}, S.~L., \& {Teukolsky}, S.~A. 1983, {Black holes, white dwarfs and neutron stars. The physics of compact objects}, \dodoi{10.1002/9783527617661}

\bibitem[{Shappee {et~al.}(2017)}]{Shappee:2017zly}
Shappee, B.~J., {et~al.} 2017, Science, 358, 1574, \dodoi{10.1126/science.aaq0186}

\bibitem[{{Shibata} \& {Hotokezaka}(2019)}]{2019ARNPS..69...41S}
{Shibata}, M., \& {Hotokezaka}, K. 2019, Annual Review of Nuclear and Particle Science, 69, 41, \dodoi{10.1146/annurev-nucl-101918-023625}

\bibitem[{Siegel \& Ciolfi(2016{\natexlab{a}})}]{Siegel:2015swa}
Siegel, D.~M., \& Ciolfi, R. 2016{\natexlab{a}}, Astrophys. J., 819, 14, \dodoi{10.3847/0004-637X/819/1/14}

\bibitem[{Siegel \& Ciolfi(2016{\natexlab{b}})}]{Siegel:2015twa}
---. 2016{\natexlab{b}}, Astrophys. J., 819, 15, \dodoi{10.3847/0004-637X/819/1/15}

\bibitem[{Sironi \& Spitkovsky(2011)}]{Sironi:2011zf}
Sironi, L., \& Spitkovsky, A. 2011, Astrophys. J., 741, 39, \dodoi{10.1088/0004-637X/741/1/39}

\bibitem[{Sironi \& Spitkovsky(2014)}]{Sironi:2014jfa}
---. 2014, Astrophys. J. Lett., 783, L21, \dodoi{10.1088/2041-8205/783/1/L21}

\bibitem[{Smartt {et~al.}(2017)}]{Smartt:2017fuw}
Smartt, S.~J., {et~al.} 2017, Nature, 551, 75, \dodoi{10.1038/nature24303}

\bibitem[{Soares-Santos {et~al.}(2017)}]{DES:2017kbs}
Soares-Santos, M., {et~al.} 2017, Astrophys. J. Lett., 848, L16, \dodoi{10.3847/2041-8213/aa9059}

\bibitem[{Sob'yanin(2016)}]{Sobyanin:2016acr}
Sob'yanin, D.~N. 2016, Astron. Lett., 42, 745, \dodoi{10.1134/S1063773716110049}

\bibitem[{Spitkovsky(2006)}]{Spitkovsky:2006np}
Spitkovsky, A. 2006, Astrophys. J. Lett., 648, L51, \dodoi{10.1086/507518}

\bibitem[{Stella {et~al.}(2005)Stella, Dall'Osso, Israel, \& Vecchio}]{Stella:2005yz}
Stella, L., Dall'Osso, S., Israel, G., \& Vecchio, A. 2005, Astrophys. J. Lett., 634, L165, \dodoi{10.1086/498685}

\bibitem[{{Stepney} \& {Guilbert}(1983)}]{1983MNRAS.204.1269S}
{Stepney}, S., \& {Guilbert}, P.~W. 1983, \mnras, 204, 1269, \dodoi{10.1093/mnras/204.4.1269}

\bibitem[{Sun {et~al.}(2019)Sun, Li, Zhang, Zhang, Bauer, Xue, \& Yuan}]{Sun:2019jaz}
Sun, H., Li, Y., Zhang, B., {et~al.} 2019, \dodoi{10.3847/1538-4357/ab4bc7}

\bibitem[{{Svensson}(1987)}]{1987MNRAS.227..403S}
{Svensson}, R. 1987, \mnras, 227, 403, \dodoi{10.1093/mnras/227.2.403}

\bibitem[{Tanaka {et~al.}(2020)Tanaka, Kato, Gaigalas, \& Kawaguchi}]{Tanaka:2019iqp}
Tanaka, M., Kato, D., Gaigalas, G., \& Kawaguchi, K. 2020, Mon. Not. Roy. Astron. Soc., 496, 1369, \dodoi{10.1093/mnras/staa1576}

\bibitem[{{Tanaka} \& {Takahara}(2010)}]{2010ApJ...715.1248T}
{Tanaka}, S.~J., \& {Takahara}, F. 2010, \apj, 715, 1248, \dodoi{10.1088/0004-637X/715/2/1248}

\bibitem[{{Timokhin} \& {Arons}(2013)}]{2013MNRAS.429...20T}
{Timokhin}, A.~N., \& {Arons}, J. 2013, \mnras, 429, 20, \dodoi{10.1093/mnras/sts298}

\bibitem[{Troja {et~al.}(2017)}]{Troja:2017nqp}
Troja, E., {et~al.} 2017, Nature, 551, 71, \dodoi{10.1038/nature24290}

\bibitem[{Utsumi {et~al.}(2017)}]{J-GEM:2017tyx}
Utsumi, Y., {et~al.} 2017, Publ. Astron. Soc. Jap., 69, 101, \dodoi{10.1093/pasj/psx118}

\bibitem[{Valenti {et~al.}(2017)Valenti, Sand, Yang, Cappellaro, Tartaglia, Corsi, Jha, Reichart, Haislip, \& Kouprianov}]{Valenti:2017ngx}
Valenti, S., Sand, D.~J., Yang, S., {et~al.} 2017, Astrophys. J. Lett., 848, L24, \dodoi{10.3847/2041-8213/aa8edf}

\bibitem[{Vurm \& Metzger(2021)}]{Vurm:2021dgo}
Vurm, I., \& Metzger, B.~D. 2021, Astrophys. J., 917, 77, \dodoi{10.3847/1538-4357/ac0826}

\bibitem[{Wang {et~al.}(2023)Wang, Beniamini, \& Giannios}]{Wang:2023qww}
Wang, H., Beniamini, P., \& Giannios, D. 2023, Mon. Not. Roy. Astron. Soc., 527, 5166, \dodoi{10.1093/mnras/stad3560}

\bibitem[{Waxman \& Bahcall(1997)}]{Waxman:1997ti}
Waxman, E., \& Bahcall, J.~N. 1997, Phys. Rev. Lett., 78, 2292, \dodoi{10.1103/PhysRevLett.78.2292}

\bibitem[{Xie {et~al.}(2022)Xie, Wei, Wang, \& Jin}]{Xie:2022igk}
Xie, L., Wei, D.-M., Wang, Y., \& Jin, Z.-P. 2022, Astrophys. J., 934, 125, \dodoi{10.3847/1538-4357/ac7c13}

\bibitem[{Yu {et~al.}(2013)Yu, Zhang, \& Gao}]{Yu:2013kra}
Yu, Y.-W., Zhang, B., \& Gao, H. 2013, Astrophys. J. Lett., 776, L40, \dodoi{10.1088/2041-8205/776/2/L40}

\end{thebibliography}
\end{document}